\documentclass[article,sort&compress]{elsarticle}

\usepackage{hyperref}
\usepackage{amsmath}
\usepackage{comment}
\usepackage{natbib}
\usepackage{subscript}
\newcommand{\tsubs}{\textsubscript}

\journal{Jou. Comp. Phys.}
{}

\begin{document}

\begin{frontmatter}

\title{A parallel interaction potential approach coupled with the immersed boundary method for fully resolved simulations of deformable interfaces and membranes}

\author[add1]{Vamsi Spandan}
\address[add1]{Physics of Fluids, University of Twente, Enschede, PO Box 217, 7500 AE, Netherlands}
\author[add2]{Valentina Meschini}
\address[add2]{Gran Sasso Science Institute, Viale Francesco Crispi, 7, L'Aquila, Italy}
\author[add3]{Rodolfo Ostilla-Monico}
\address[add3]{Harvard John A. Paulson School of Engineering and Applied Sciences, Harvard University, Cambridge, MA 02138, USA}
\author[add1,add6]{Detlef Lohse}
\address[add6]{Max Planck Institute for Dynamics and Self-Organization, 37077, G\"ottingen, Germany}
\author[add5]{Giorgio Querzoli}
\address[add5]{DICCAR - University of Cagliari, via Marengo 2, 09123 Cagliari, Italy}
\author[add4]{Marco D de Tullio}
\address[add4]{Politecnico di Bari, Via Re David 200, 70125, Bari, Italy}
\author[add1,add7]{Roberto Verzicco}
\address[add7]{University of Rome 'Tor Vergata', Via del Politecnico, Rome 00133, Italy}


\begin{abstract}
In this paper we show and discuss the use of a versatile interaction potential approach coupled with an immersed boundary method to simulate a variety of flows involving deformable bodies. In particular, we focus on two kinds of problems, namely (i) deformation of liquid-liquid interfaces and (ii) flow in the left ventricle of the heart with either a mechanical or a natural valve. Both examples have in common the two-way interaction of the flow with a deformable interface or a membrane. The interaction potential approach (de Tullio \& Pascazio, \emph{Jou. Comp. Phys.}, 2016; Tanaka, Wada and Nakamura, Computational Biomechanics, 2016) with minor modifications can be used to capture the deformation dynamics in both classes of problems. We show that the approach can be used to replicate the deformation dynamics of liquid-liquid interfaces through the use of ad-hoc elastic constants. The results from our simulations agree very well with previous studies on the deformation of drops in standard flow configurations such as deforming drop in a shear flow or a cross flow. We show that the same potential approach can also be used to study the flow in the left ventricle of the heart. The flow imposed into the ventricle interacts dynamically with the mitral valve (mechanical or natural) and the ventricle which are simulated using the same model. Results from these simulations are compared with ad-hoc in-house experimental measurements. Finally, a parallelisation scheme is presented, as parallelisation is unavoidable when studying large scale problems involving several thousands of simultaneously deforming bodies on hundreds of distributed memory computing processors.
\end{abstract}

\begin{keyword}
\texttt{elsarticle.cls}\sep \LaTeX\sep Elsevier \sep template
\MSC[2010] 00-01\sep  99-00
\end{keyword}

\end{frontmatter}

\section{Introduction}
The interaction between fluid flow and an immersed elastic body (fluid or solid) has been studied extensively over the last few decades due to its wide range of applications, for example, bubbles and drops dispersed in a turbulent flow \citep{balachandar2010turbulent,tryggvason2013multiscale}, red blood cells flowing through blood vessels \citep{freund2014numerical}, pumping motion of ventricles and valves in the heart \citep{sotiropoulos2016fluid,mittal2016computational}, oscillation of large structures such as aircraft wings and high-rise buildings \citep{dowell2004modern}, etc. While the source of elasticity of the immersed body in each of these phenomena is different, the interplay between a deformable immersed body and a surrounding inhomogeneous time dependent flow can result in a complex non-linear system where they determine each other's behaviour in a coupled manner. Additionally, the presence of multiple bodies with different static and dynamic properties interacting with each other gives rise to a wide range of control parameters which makes these systems extremely challenging to study. Over the last few decades, tremendous amount of effort has been devoted to the modelling and simulation of such systems which are often classified in literature as Fluid-Structure Interaction (FSI) problems. Among a variety of techniques developed to tackle FSI problems, the immersed boundary method (IBM) \cite{peskin1972flow,peskin2002immersed,mittal2005immersed} has gained immense popularity and has been instrumental in making efficient and accurate simulations of several complex flow systems such as cardiac and vascular hemodynamics \citep{sotiropoulos2016fluid,mittal2016computational}, suspensions of rigid spheres \cite{uhlmann2014sedimentation,picano2015turbulent,prosperetti2015life,
fornari2016sedimentation}, deformable bubbles or drops \citep{schwarz2016immersed}, vehicle aerodynamics \citep{iaccarino2003immersed,de2011recent} etc. possible.

One of the biggest advantages of the IBM is that it relies on the use of a single underlying mesh for the fluid flow (hereafter referred to as Eulerian mesh) which does not have to conform/adapt with the moving/deforming immersed body \citep{peskin1972flow,peskin2002immersed,mittal2005immersed}. This eliminates the complex and computationally expensive procedure of Eulerian mesh regeneration every time step as the immersed body moves or deforms, resulting in the decoupling of the mesh required for the flow solver from the position and morphology of the immersed body. The surface of the immersed body is discretised independently of the Eulerian mesh and is often called a Lagrangian or a  structural mesh. The influence of the immersed body on the flow can be achieved through a volume averaged body force in the fluid governing equations after a careful transfer of information between the Eulerian and Lagrangian meshes. Additionally, the time invariant nature of the Eulerian mesh makes IBM promising for parallelisation on multiple distributed memory computing processors and this has led to breakthroughs in simulations of highly turbulent flows around complex geometries.

In this paper, we build upon the work of de Tullio and Pascazio \cite{detullio2016moving} and describe the coupling of a multi-physics interaction potential approach with a finite-difference Navier-Stokes solver which can be used to simulate a variety of fluid-structure interaction problems with different sources of elasticity. In particular, we focus on two problems which have been the focus of a large part of the fluid-dynamics community for several years, namely (i) deformable fluid-fluid interfaces (drops or bubbles) in a given flow and (ii) flow dynamics in the left ventricle of the human heart with either a mechanical or natural mitral valve.

Understanding the behaviour of dispersed particles, bubbles or drops in a turbulent flow is a standalone field in itself and several techniques have been developed explicitly to tackle such problems (e.g. point-particle, Volume of Fluid, level-set, front tracking, Physalis etc.) \citep{prosperetti2007computational}. The complexity in the simulation of such flows arises from the wide range of length and time scales and the regimes involved, see \citep{crowe1996numerical,magnaudet2000motion,tryggvason2013multiscale} for detailed reviews. When the dispersed phase is rigid, smaller than the smallest length scale (Kolmogorov scale) in the flow and can be described using simple shapes (spherical or ellipsoidal), the drop/bubble momentum equations can be simplified into computationally inexpensive force balances which rely on empirical correlations \citep{balachandar2010turbulent}. However, when the bubbles or drops become larger and experience inhomogeneous flow conditions over their surfaces they can deform into complex shapes while strongly interacting with the fluid. In such cases the singularity of the surface tension term in the governing equations gives rise to several numerical hurdles \citep{scardovelli1999direct}. The complex algorithms and procedures put in place to tackle these numerical instabilities have restricted the scale of multiphase flows that can be studied, for example, state-of-art parallel simulations can only reach up to $O(10^2)$ deformable drops/bubbles in a reasonably turbulent flow \cite{tryggvason2013multiscale}.

The emergence of IBM in simulating turbulent multiphase flows began with the seminal work of Uhlmann \cite{uhlmann2005immersed}. He proposed an alternative direct forcing scheme which requires computing the IBM forcing term on Lagrangian markers uniformly distributed over the surface of the immersed body. Several additional features and improvements to this initial idea has led to a massive growth in the use of IBM for dispersed multiphase flows. Breugem \cite{breugem2012second} built upon the work of Uhlmann \cite{uhlmann2005immersed} and used a multi-direct forcing scheme for a better approximation of the interfacial no-slip boundary condition. Vanella and Balaras \cite{vanella2009moving} used a relatively costlier moving-least-squares (MLS) interpolation to build transfer functions between the Eulerian and Lagrangian meshes. More recently, Schwarz {\it et al.} \cite{schwarz2015temporal} proposed a \lq virtual mass\rq\ approach to overcome the numerical instabilities arising from added mass effects of dispersed light bodies. The same group also demonstrated the use of spherical harmonics to simulate bubbles or drops of varying shape which is computed through a minimization of the local displacement energy induced by the pressure and surface tension forces \cite{schwarz2016immersed}. To simulate deformable bodies with several degrees of freedom, de Tullio and Pascazio \cite{detullio2016moving} employ a simple spring network model with an interaction potential along with a Navier-Stokes solver. They demonstrated that numerical simulations of a variety of problems involving large accelerations can be realised using this strongly coupled interaction potential Navier-Stokes IBM solver (e.g. motion of rigid bodies, thin elastic structures, flapping or flexible bodies like flags or leaflets of bio-prosthetic aortic valve etc.).

In the first part of this paper, we extend the approach used by de Tullio and Pascazio \cite{detullio2016moving} to be able to simulate closed liquid-liquid interfaces such as drops/bubbles under certain flow conditions. Given the complexity and numerical challenges of more advanced methods (for e.g. V.O.F, level-set, front-tracking), we show that reasonably accurate simulations of flows involving deformable interfaces can be achieved through this simple multi-physics approach. The interfacial boundary condition involving liquid-liquid interfaces is taken to be no-slip given its straight-forward implementation. In most of real-life situations drops or bubbles dispersed in a liquid are contaminated either through surfactants or impurities which leads to a no-slip interfacial boundary condition. The IBM algorithm can also be modified accordingly to account for a stress-free interfacial boundary condition as shown by Kempe {\it et al.} \cite{kempe2015imposing}.

IBM has also emerged as a front runner in the field of cardio-vascular hemodynamics as has been evidenced in several recent studies \cite{choi2015new,seo2013multiphysics,de2009direct,de2011fluid,zheng2012computational,seo2013effect,seo2014effect,vedula2014computational}. A main impulse in developing this area of study from a computational fluid dynamics point of view is the increasing demand from the medical community for scientifically rigorous and quantitative investigations of cardiovascular diseases. Detailed computational studies can assist surgeons in understanding how various surgical solutions can affect blood circulation and guide the choice of the most appropriate procedure for a specific patient or type of patients. Moreover, they are of course non-invasive in contrast to \emph{in vivo} investigation and potentially yield more complete and detailed information than \emph{in vitro} experiments. The major bottleneck in conducting fully resolved simulations of the complete human heart is created by the complex deformation dynamics of the various ventricles and valves which interact with the pulsatile blood flow (see \citep{mittal2016computational} for a recent review). Various approaches have been employed over the years to achieve realistic and reliable cardiac hemodynamic simulations. One approach makes use of medical imaging techniques such as 4D cardiac tomography (CT)/Echo and cardiac magnetic resonance (CMR) imaging to reconstruct a patient-specific time dependent kinematic model of the heart walls. An alternative approach is to use available models of the heart functionality along with the biophysical component of cardiac electromechanics. This approach has been used in the \lq Living Heart Project\rq\ \citep{baillargeon2014living}, where a fully coupled electro-mechanic and hemodynamic simulation is realised, and more recently by Choi \emph{et al.} \citep{choi2015new}, who coupled a multi scale model for electromechanics with the Navier-Stokes equations for the flow dynamics. The work of Zheng {\it et al.} \cite{zheng2012computational} and Seo and Mittal \cite{seo2013effect} focussed on intra-ventricular flow and the accompanying pathologies under the effect of a diastolic flow pattern. In these simulations, the left ventricle was deformed with a prescribed motion derived from various imaging techniques. A similar hemodynamic model for the left ventricle is used by Seo \emph{et al.} \cite{seo2014effect} to analyse the effect of mitral valve on the flow dynamics where the motion of the mitral valve is imposed through a kinematic model but not with a fully coupled fluid-structure interaction simulation.

The motivation in employing kinematic models to describe the motion of ventricles and valves instead of a fully coupled FSI simulation is to eliminate the massive computational cost in solving the three dimensional Cauchy-Navier equations for the immersed elastic body along with the Navier-Stokes equations. Although numerical simulations with pre-defined ventricle/valve motion is a challenging task in itself, in reality the motion of the ventricle, valves, and the fluid are coupled to each other and can govern each other's motion. 

In the second part of this paper, we show that a full fluid-structure interaction simulation of a ventricle with a mechanical or biological valve can be made possible through the use of the interaction potential approach for elastic deformation. In particular, we focus on the simulation of the left ventricle of the human heart along with a physical mitral valve in both pathological and physiological conditions.

In addition to having a versatile FSI approach which can simulate a variety of elastic interfaces or membranes such as bubbles, drop, ventricles, valves etc., it is also necessary to ensure that the method is not computationally too demanding. The full FSI simulation of the left ventricle with a physical valve which is described in detail later is performed within the equivalent of 48 CPU hours on a single processor with an Eulerian grid of 150x150x150 and a Lagrangian mesh on the ventricle with approximately 50000 triangular elements which shows the computationally inexpensive nature of the interaction potential approach. Additionally, these simulations were compared and validated with in-house ad-hoc laboratory experiments to ensure the reliability of the results.

However, the IBM described in this paper makes use of MLS interpolations whose computational cost increases rapidly with increase in the resolution of the immersed bodies. New algorithms or parallel implementation of the computation on distributed memory processors thus become an invaluable tool in scaling up fully resolved flows interacting with multiple moving/deforming immersed bodies. In particular, parallelisation is an attractive prospect given the increasing availability of cost-efficient high performance computing facilities.

A parallel implementation of a flow solver involving multiple deforming bodies is not a straightforward task due to many algorithmic complexities. The challenge lies underneath the fact that two different meshes (Eulerian and Lagrangian) are required for the complete solution and different parallelisation strategies would be required to make use of multiple processors effectively. To understand this, we briefly describe the various steps that are need to be completed on both the Eulerian and Lagrangian mesh to fully simulate a FSI problem: (i) Computing the solution for the fluid phase governing equations on the Eulerian mesh. (ii) Interpolating the Eulerian flow velocity on the Lagrangian markers and computing the required volume averaged IBM forcing term. (iii) Enforcing the interfacial boundary condition on the Eulerian mesh through the volume averaged forcing term in the fluid governing equations. (iv) Transferring the local flow conditions onto the Lagrangian mesh followed by transporting and deforming the immersed body. While the Eulerian mesh is fixed in space and time, the Lagrangian mesh can move with multiple degrees of freedom and be distributed over several processors. It is also important to note that unlike the Eulerian mesh, the connectivity of the Lagrangian mesh needs to be stored by all processors. Additionally, the transfer functions built to exchange information between the Eulerian and Lagrangian mesh may require data from neighbouring processors, thus requiring the storage of multiple ghost layers (information from neighbouring processors) in the processor's memory. 

Given all the above challenges, devising a simple and effective parallelisation algorithm can be highly non-trivial. Keeping this in mind, in the last part of this paper we describe a parallelisation scheme designed to track the time evolution of several deformable bodies (e.g. vesicles, drops, biological tissues etc.) immersed in turbulent flows. This strategy is built upon the already underlying parallelisation scheme implemented for the fluid solver, thus reducing the downtime of overall code development. In particular, the benefits of the parallelisation is oriented towards simulation of dispersed phase systems with several thousand deforming drops, bubbles, vesicles or bodies moving in a highly turbulent carrier fluid phase.

In the next section we give a brief overview of the governing equations for the solution of the fluid phase, implementation of the immersed boundary method using Moving Least Squares (MLS) and the interaction potential approach for computing the deformation of elastic bodies. In section 3, we show how the interaction potential approach can be used to study deformation of drops/bubbles where the flow is dynamically coupled with the interface morphology. These results are validated with analytical solutions and experimental measurements taken from literature. In section 4, we describe the simulation of the full left ventricle with both mechanical and natural mitral valves in addition to comparing the results from our simulations with ad-hoc in-house experiments. In section 5, we discuss the data structures required and also the parallelisation strategy to scale up the problem to study several thousand deforming immersed bodies. Finally, we provide a summary and outlook in section 6.

\section{Governing equations and numerical scheme}
\label{sec:eqn}
\subsection{Fluid phase}
For the fluid phase we solve the incompressible Navier-Stokes equations in a Cartesian box as given in equations (\ref{eqn:ns}), (\ref{eqn:con}).

\begin{equation}
\frac{\partial \textbf u}{\partial t}+\textbf u \cdot \nabla \textbf u=-\nabla p+\frac{1}{Re}\nabla^2 \textbf u +\textbf f_b,
\label{eqn:ns}
\end{equation}
\begin{equation}
\nabla \cdot \textbf u=0 .
\label{eqn:con}
\end{equation}

The Reynolds number of the flow is defined based on a characteristic length scale $L$ and velocity scale $U$ as $Re=UL/\nu$; $\nu$ is the kinematic viscosity of the fluid. $\bf u$, $p$ are the velocities and pressure in the flow while $\bf f_b$ is the volume averaged force arising from the IBM and is included to enforce the interfacial boundary condition.
A conservative second-order centred finite-difference scheme with velocities on a staggered grid is used for spatial discretisation; explicit Adams-Bashforth scheme is used to discretise the non-linear terms while an implicit Crank-Nicholson scheme is used for the viscous terms. Treating all the viscous terms implicitly results in a large sparse matrix which is avoided by an approximate factorisation of the sparse matrix into a tridiagonal matrix. Time integration is performed via a self starting fractional step third order Runge-Kutta (RK3) scheme. The pressure required to enforce mass conservation is computed by solving a Poisson equation for a pressure correction. The code for single phase flows has already been tested extensively in previous studies for a variety of flow configurations and additional details of the numerical scheme can be found in \cite{verzicco1996finite,van2015pencil}.

\subsection{Dispersed phase: Immersed Boundary Method}

\begin{figure}
  \centerline{\includegraphics[scale=1.0]{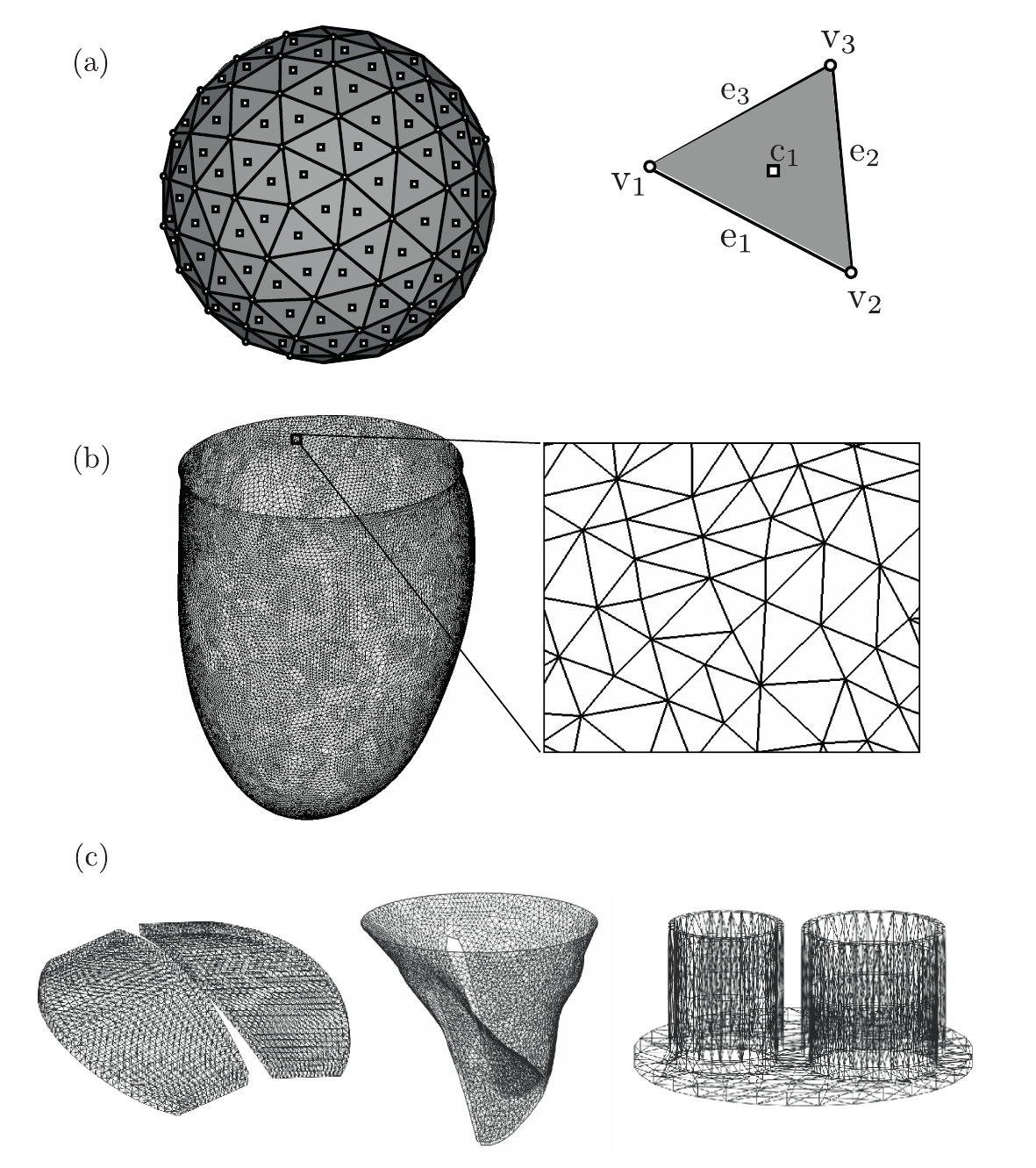}}
  \caption{Schematic of the Lagrangian mesh (a) A sphere discretised using triangular elements. On the right a single triangular element is decomposed into three vertices v\tsubs{1},v\tsubs{2},v\tsubs{3} (circles), three edges e\tsubs{1},e\tsubs{2},e\tsubs{3} and one centroid c\tsubs{1} (square). (b) Full structure of the left ventricle with a zoomed-in area showing the triangulated network. (c) Rest of the components of the full left ventricle structure. The left panel shows the leaflets of prosthetic mechanical mitral valve, the middle panel shows the natural mitral valve and the right panel the channels for mitral and aortic valves.}
\label{fig:lagmesh}
\end{figure}

We now describe the procedure of constructing the Lagrangian mesh and the schemes used to transfer flow quantities between the Lagrangian and Eulerian mesh which is a crucial ingredient in IBM. In figure \ref{fig:lagmesh}, we show a schematic of the various Lagrangian meshes used in this study. Any given surface (closed or open) is discretised into triangular elements where each element is composed of three vertices (v\tsubs{1}, v\tsubs{2}, v\tsubs{3}) which are connected by edges (e\tsubs{1}, e\tsubs{2}, e\tsubs{3}). The position of the centroid (c\tsubs{1}) of each triangular element is computed based on the co-ordinates of the vertices, the mass being uniformly distributed on the single triangular element. Figure \ref{fig:lagmesh}(a) shows a sphere discretised into triangular elements along with a schematic showing the composition of each triangle. In figure \ref{fig:lagmesh}(b), we show the discretised geometry of the left ventricle and figure \ref{fig:lagmesh}(c) shows the remaining auxiliary components.

Following the idea of Uhlmann \cite{uhlmann2005immersed} the force required to enforce the interfacial boundary condition is first computed on markers laid out on the Lagrangian mesh and then transferred to the Eulerian mesh. Here, we consider the triangle centroids to be the Lagrangian markers and are responsible for enforcing the interfacial boundary condition. The vertices and edges of the discretised triangular elements play a role in the deformation dynamics and will be explained later. The next step is to build a transfer function around each Lagrangian marker (here the centroid c\tsubs{i}) which would be used to exchange information between the Eulerian and Lagrangian mesh. We adopt the Moving Least Squares (MLS) \cite{lancaster1981surfaces,liu2005introduction} approach which is part of the class of meshless approximations and has been used previously in several fields such as element free Galerkin methods \cite{belytschko1994fracture,belytschko1996meshless,krongauz1996enforcement,hegen1996element,atluri1999analysis}, computer graphics \cite{schaefer2006image,fleishman2005robust,kolluri2008provably,zeng2004curve,kobbelt2004survey}, and also recently for IBM \cite{vanella2009moving,detullio2016moving}. In order to compute this transfer function we first need to build a support domain centred around each Lagrangian marker which consists of all Eulerian grid nodes closer than a threshold value in each direction. By taking a threshold value of $1.5\Delta x_i$ in each direction, a three-dimensional support domain consisting of $N_e=27$ (3x3x3) Eulerian nodes is built around each Lagrangian marker. The next step is to use these Eulerian nodes and build a transfer function through which any quantity $q_i$ defined on the Eulerian nodes can be interpolated on the Lagrangian marker (c\tsubs{i}). The same transfer function can be used to extrapolate the force computed on the Lagrangian markers ($F_i$) to the Eulerian mesh ($f_i$).

The MLS interpolation of $q_i$ at a Lagrangian marker (c\tsubs{i}) is defined as follows.

\begin{equation}
Q_i(\pmb x) = \pmb p^T(\pmb x)\pmb a(\pmb x)=\sum_{j=1}^{m} p_j(\pmb x)a_j(\pmb x)
\label{eqn:mls}
\end{equation}
In equation (\ref{eqn:mls}), $Q_i$ is the quantity interpolated on the Lagrangian marker while $\pmb p^T(\pmb x)$ is a basis function vector with dimension $m$. $\pmb a(\pmb x)$ is the vector of coefficients obtained by minimising the weighted L2 norm and $\pmb x$ is the position vector of the Lagrangian marker. In this work we consider a linear basis function with $\pmb p^T(\pmb x)=[1,x,y,z]$, i.e. $m=4$, which is cost-efficient and also able to represent the gradients in the Eulerian field with second order accuracy.

\begin{equation}
J=\sum_{k=1}^{N_e}W(\pmb x-\pmb x^k)[\pmb p^T(\pmb x^k)\pmb a(\pmb x)-q_i^k]^2
\label{eqn:mls_j}
\end{equation}
Here $W(\pmb x-\pmb x^k)$ is a weight function; we use the exponential weight function which is given as follows.

\begin{equation}
W(\pmb x-\pmb x^k)=
\begin{cases}
    e^{-(r_k/\alpha)^2},& r_k\leq 1\\
    0,              & r_k > 1
\end{cases}
\label{eqn:mls_w}
\end{equation}
where $\alpha$ is a constant of shape parameter and $r_k$ is given by
\begin{equation}
r_k=\frac{|\pmb x-\pmb x^k|}{r_i}
\label{eqn:mls_rk}
\end{equation}
$r_i$ is the size of the support domain in the $i^{\text{th}}$ direction. Other commonly used shape functions are the cubic spline and quadratic spline functions and a spline function with any order of continuity can be constructed using the steps detailed in Liu and Gao \cite{liu2005introduction}. Minimising $J$ in equation (\ref{eqn:mls_j}) leads to $\pmb A(\pmb x)\pmb a(\pmb x)=\pmb B(\pmb x)\pmb q_i^k$ where
\begin{equation}
\pmb A(\pmb x)=\sum_{k=1}^{N_e}W(\pmb x-\pmb x^k)\pmb p(\pmb x)\pmb p^T(\pmb x^k)
\label{eqn:mls_A}
\end{equation}
\begin{equation}
\pmb B(\pmb x)=[W(\pmb x-\pmb x^1)\pmb p^T(\pmb x^1)...W(\pmb x-\pmb x^{N_e})\pmb p^T(\pmb x^{N_e})]
\label{eqn:mls_B}
\end{equation}
\begin{equation}
\pmb q_i=[q_i^1...q_i^{N_e}]^T
\label{eqn:mls_q}
\end{equation}
Combining all the above equations, the interpolated quantity $Q_i$ can be expressed as follows.
\begin{equation}
Q_i(\pmb x)=\pmb \phi^T(\pmb x)\pmb q_i=\sum_{k=1}^{N_e}\phi_k^l(\pmb x) q_i^k
\label{eqn:mls_Q}
\end{equation}
$\pmb \phi^T(\pmb x)=\pmb p^T(\pmb x)\pmb A^{-1}(\pmb x)B(\pmb x)$ is the transfer function containing the shape function coefficients for each Lagrangian marker. This shape function is used to interpolate the value of the intermediate Eulerian velocity $\hat u_i$ at the exact location of all Lagrangian markers and the volume force in each direction is calculated as $F_i=(V^b_i-U_i)/\Delta t$, where $V^b_i$ is the desired velocity boundary condition on the Lagrangian marker (c\textsubscript{i}) and $U_i$ is the Eulerian flow velocity interpolated on the Lagrangian marker using MLS. This force needs to be transferred back to the Eulerian mesh using the same transfer function built for interpolation in equation (\ref{eqn:mls_Q}) under the constraint that the total force is conserved during the extrapolation. This gives the Eulerian force as $f_{b,i}^k=\sum_{k=1}^{N_l}c_l\phi_k^lF_i^l$, where $N_l$ is the number of Lagrangian markers associated with a Eulerian point $k$. $c_l$ is a scaling factor obtained by imposing the condition that there is no net-gain/loss in the IBM force while transferring flow information from the Lagrangian mesh to the Eulerian mesh which results in the following.

\begin{equation}
c_l=\frac{\Delta V^l}{\sum_{k=1}^{N_e}\phi_k^l\Delta V^K}
\label{eqn:mls_cl}
\end{equation}
where $\Delta V^l$ is the forcing volume associated with each Lagrangian marker and is computed as $\Delta V^l=A_lh_l$. $A_l$ is the area of the triangular element associated with the Lagrangian marker (area composed by v\tsubs{1}, v\tsubs{2}, v\tsubs{3} in figure \ref{fig:lagmesh}) and $h_l=1/3\sum_{k=1}^{N_e}\phi_k^l(\Delta x^k+\Delta y^k+\Delta z^k)$. Here it is important to note that the transfer functions built using this approach conserves momentum on both uniform and stretched grids while reasonable accuracy is retained for torque equivalence on slightly stretched grids \cite{vanella2009moving,detullio2016moving}. For example, Vanella and Balaras \cite{vanella2009moving} report that with 10 \% grid stretching, the net loss/gain in torque conservation is less than 0.5 \%.

The calculation of hydrodynamic forces (pressure and viscous stresses) acting on the surface of any dispersed body in an IBM simulation is not straightforward as the Lagrangian and the Eulerian meshes do not necessarily align with each other at a given time instant. Since the surface of the dispersed bodies are discretised using triangular elements, the local pressure and viscous forces are first computed on the Lagrangian markers (centroids of triangular elements); the total external force on a triangular element $l$ with area $A_l$ and surface normal $\pmb n_l$ is calculated as $\pmb F_\text{ext}^l=(-p_l\pmb n_l + \pmb \tau_l\cdot \pmb n_l)A_l$. To evaluate $p_l$ and $\tau_l$, which are the pressure and viscous forces acting on a triangular element $l$, respectively a probe is sent along the normal of each triangular element with its centroid as the origin. The length of the probe $h_l$ is equal to the mean local grid size and the MLS interpolation described above is used to interpolate both pressure and velocity gradients at the end point of the probe. The pressure on the Lagrangian marker (centroid) is computed as $p_l=p_l^*+h_l\frac{D\pmb u_l}{Dt}\cdot \pmb n_l$; $p_l^*$ is the pressure at the probe endpoint and $\frac{D\pmb u_l}{Dt}$ is the acceleration of the Lagrangian marker \cite{yang2006embedded,vanella2009moving,detullio2016moving}. The shear stress $\pmb \tau_l$ on the Lagrangian marker is computed based on the velocity gradients interpolated at the probe endpoint. This holds true under the assumption that the velocity of the fluid near the surface of the body varies linearly. An important note here is that when the immersed body is an open surface such as a flag, ventricle or a valve, the pressure and viscous forces need to computed on both sides of the surface thus requiring two probes sent along the normal to every triangular element, one each along the positive and negative normal, respectively i.e. $\pmb F_\text{ext}^l=[(-(p_l^+-p_l^-)\pmb n_l^+ + (\pmb \tau_l^+-\pmb \tau_l^-)\cdot \pmb n_l^+)]A_l$; the subscripts $^+$ and $-$ represent quantities evaluated on the end points of the probe on either side of the surface. Additional details on this can be found in \cite{detullio2016moving,yang2006embedded,vanella2009moving}.

\subsection{Interaction potential approach for deformation}
As mentioned in section 1, the dynamics of deformation is computed based on a minimum energy concept which we describe here briefly (see \cite{detullio2016moving} for more details). The surface of any immersed body is first discretised using triangular elements (c.f. figure \ref{fig:lagmesh}) the edges of which are composed of hypothetical linear/non-linear springs thus resulting in a two-dimensional network of springs. The mass of the immersed body is assumed to be uniformly distributed among the vertices of the triangular elements, i.e. the mass of each triangular node $m=A_sh\rho_m/N_\text{vertices}$, where $A_s$ is the total surface area of the elastic membrane, $h$ is its thickness, $\rho_m$ is the density of the membrane and $N_\text{vertices}$ are the total number of vertices in the spring network. Under the influence of external forces such as pressure fluctuations or viscous stresses, the spring network undergoes deformation thus storing potential energy into the system. The potential is converted to nodal forces acting on individual triangular vertices through a spatial derivative operation. The acceleration of each triangle vertex is computed from the force and integrated based on Newton's second law of motion; thus the position of each individual vertex is updated independently of which immersed body it belongs to. 

The first form of potential is the in-plane elastic potential ($W_e$) which comes from the work done by an external force parallel to the plane of a triangular face and is converted into elastic energy stored into every spring connecting the triangle. We also consider an out-of-plane deformation ($W_b$) for which the total potential is computed based on a bending spring connecting the centroids of two adjacent triangular faces. This out-of-plane bending potential is stored in a pair of two faces sharing an edge and is a function of the contact angle between them. Additional potentials can be included which constrain the geometrical properties of the overall immersed body. For example, we can include a volume or area potential ($W_v$ or $W_a$) which is a function of the change in volume/area of a single element with respect to an initial reference state. All the individual potentials are calculated as given in the equations below.

\begin{equation}
W_e=\frac{1}{2}k_e\pmb x^2
\label{eqn:elaspot}
\end{equation}
\begin{equation}
W_b=k_b(1-cos\theta)
\end{equation}
\begin{equation}
W_v=\frac{1}{2}k_v\Bigg(\frac{V-V_0}{V_0}\Bigg)^2 V_0
\end{equation}
\begin{equation}
W_a=\frac{1}{2}k_a\Bigg(\frac{A-A_0}{A_0}\Bigg)^2 A_0
\label{eqn:areapot}
\end{equation}
In the above equations, $k_e$, $k_b$, $k_v$, and $k_a$ are the elastic constants for in-plane deformation, out-of-plane deformation, volume constraint and area constraint potentials, respectively. $\pmb x$ is the change in length of a single edge; $\theta$ is the angle between the normals of two triangular faces sharing an edge; $V_0$, $V$ and $A_0$, $A$ are the corresponding initial (reference) and deformed volumes and areas of each triangular element, respectively. While the equations (\ref{eqn:elaspot})-(\ref{eqn:areapot}) can be used to simulate homogeneous isotropic materials, the interaction potential approach can also used for inhomogeneous anisotropic materials by changing the functional form of the elastic potentials \cite{detullio2016moving}. 

Once the forces on each of the triangular node (vertex) is known, individual nodes are moved based on the equation $m\ddot{\pmb x}^{\text v_i} = F_\text{ext}^{\text v_i}+F_\text{int}^{\text v_i}$; $F_\text{ext}^{\text v_i}$ and $F_\text{int}^{\text v_i}$ are the external and internal forces acting on the triangular nodes ($\text v_i$), $\ddot{\pmb x}$ and $m$ are the acceleration and mass of individual nodes. In the previous section $F_\text{ext}$ was calculated on the centroid of each triangle. This force is transferred to an individual triangular vertex as $F_\text{ext}^{\text v_i}={\sum_{j=1}^{\text n_{fi}}}(1/3)F_\text{ext}^{\text c_j}$; where $\text n_{fi}$ is the number of faces each vertex is connected to and $F_\text{ext}^{\text c_j}$ is the external force computed on the triangle centroid (Lagrangian marker) $\text c_j$. As described previously, the calculation of $m$ which is the mass of individual triangular node is straightforward for surfaces made of materials where the density and thickness is known a priori. In cases where the immersed bodies are drops or bubbles, calculating $m$ of the triangular nodes becomes tricky as there is no physical definition of the density and thickness of a liquid-liquid interface. In such a case $m$ of the triangular nodes becomes a free parameter and to overcome this we fix the value of $m=1$ and then correspondingly tune the elastic constants. This is detailed more in the next section. Computing the individual potentials according to equations (\ref{eqn:elaspot})-(\ref{eqn:areapot}) in the interaction potential approach requires selection of several parameters ($k_e$, $k_b$, $k_v$, $k_a$). Once again, this step is straightforward for membranes where the elastic moduli are already known \cite{detullio2016moving}. It is important to note that when the surface is discretised with non-uniform triangles such that the lengths of the edges of triangle vary, $k_e$ should be computed based on the model proposed by van Gelder \cite{gelder1998approximate}. Simulating liquid-liquid interfaces using the interaction potential approach would require the use of ad-hoc elastic constants as again there is no direct physical correlation between the elastic constants and the surface properties of a liquid-liquid interface. The procedure of estimating these ad-hoc elastic constants will be described in detail in the next section.

Here it is important to note that modelling an elastic membrane or an interface using the interaction potential approach is a discrete formulation of the elasticity governing equations and thus a simplification of the existing continuum models. Such a formulation is useful and necessary for complex simulations of several immersed deformable bodies owing to its simplicity and lower computational cost. It has been shown in previous studies that through a careful design of the spring network and the selection of appropriate elastic constants the mechanics of several elastic membranes can be exactly reproduced \cite{detullio2016moving,chen2014investigation,fedosov2010systematic}. Here we would like to note that the derivation of the interaction potential approach is not unique to our work and variants of this method have already been used previously to predominantly study red blood cells \cite{fedosov2010systematic,chen2014investigation,dupin2008lattice,kruger2016effect}. In this work we show that this approach can be used for large scale flows with dispersed drops/bubbles and also biological membranes with full fluid-structure interaction, for example flow in heart ventricles with valves. 

Another important issue in the simulations involving FSI problems is the type of coupling used i.e. loose (explicit) versus strong (implicit). In the loosely coupled (explicit) case, the fluid and the immersed body governing equations are solved separately one after the another with a transfer of information between them every time step. On the other hand, in the strongly coupled (implicit) case the governing equations are solved in an iterative manner for each time step using a predictor corrector scheme until sufficient convergence is achieved. A detailed solution procedure for a strongly coupled IBM-FSI Navier Stokes solver with the provision for the interaction potential approach is given in de Tullio and Pascazio \cite{detullio2016moving} where the governing equations are solved using a Hamming\'s fourth order predictor-corrector scheme. In our code, we have provisions for both a strong (implicit) and weak (explicit) coupling between the fluid and the immersed body. For the simulations shown in the following sections, loose coupling is used given its computationally inexpensive nature. Also it has to be remembered that strong coupling is only needed when added mass effects from the immersed body become important and the recent work of Schwarz {\it et al.} \cite{schwarz2015temporal} gives insights into how this can be tackled smartly while there is still loose coupling between the fluid and the immersed body. We now move on to combining DNS of the fluid governing equations along with a moving least squares IBM coupled with the interaction potential approach to simulate deformable drops/bubbles and heart ventricles/valves.    

\section{Liquid-liquid interface dynamics using the potential approach}
In order to simulate the interfacial behaviour of drops or bubbles using the interaction potential approach (hereafter called the IP model), we first need to devise a method to compute the elastic constants of a given spring network which can represent a liquid-liquid interface with a given surface tension. As mentioned earlier, this is not straight-forward since there is no direct physical correlation between the elastic constants and the surface tension of a liquid-liquid interface. Here we use a reverse-engineered approach and perform a single simulation with a set of intuitively chosen elastic constants and estimate the surface tension of the immersed drop by comparing its morphology with previously known analytical solutions. By using the same set of elastic constants but for different flow conditions we also show that such an approach is self-consistent and reliable. Our goal here is to show that the IP model for deformation can be reliably used to replicate the deformation dynamics of liquid-liquid interfaces under given flow conditions.

\subsection{Deformation of a neutrally buoyant drop in shear flow}
 For the first test case we choose the system of a neutrally buoyant drop deforming in a laminar shear flow which has a simple configuration and a limited set of control parameters. Variants of this problem have been studied for a long time and several analytical and phenomenological models already exist in literature which can accurately predict the deformation dynamics of the immersed drop \cite{taylor1932viscosity,taylor1934formation,maffettone1998equation}. The deformation of an initially spherical drop immersed in a shear flow arises from competing actions of the viscous drag which tends to stretch the drop and the surface tension forces acting to recover the initial spherical shape. For this simulation, we use a Cartesian box which is wall-bounded in the vertical direction ($\hat e_z$) and fully periodic in the horizontal directions. The top and bottom walls move in the opposite direction parallel to each other with the same velocity to generate a laminar shear in the domain. A triangulated sphere as shown in figure \ref{fig:lagmesh}(a) is positioned in the flow at a distance $0.5L_z$ from the walls ($L_z$ is the gap between the walls).

The degree of deformation and orientation of a viscous drop in the presence of a velocity gradient depends on the Capillary number $Ca=\mu_fR\dot \gamma/\sigma$, where $R$, $\dot \gamma$, $\sigma$ and $\mu_f$ are the drop radius, local shear rate, surface tension and fluid viscosity, respectively. For these simulations the viscosity ratio of the droplet and the carrier phase is set to 1, i.e. $\hat \mu=\mu_d/\mu_c=1$. Since this simulation will be used to \lq tune\rq\ the elastic constants to replicate a liquid-liquid interface, the IBM forcing $\pmb f$ in equation (\ref{eqn:ns}) is set to zero. This is done so that the immersed sphere only experiences the forces generated due to the laminar shear from the moving walls and not due to any boundary layer formation on the sphere surface.

\begin{figure}
  \centerline{\includegraphics[scale=0.99]{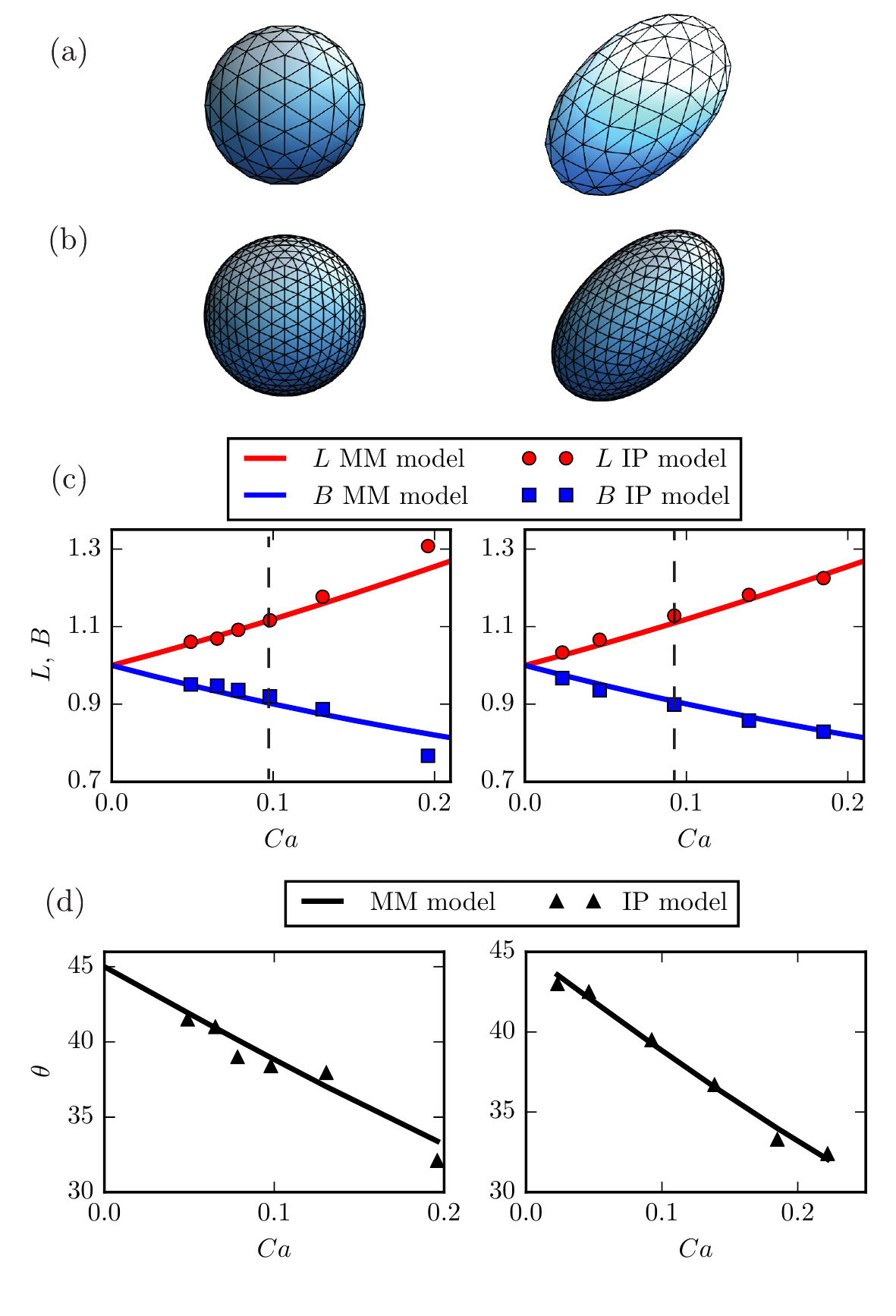}}
  \caption{Deformation of a neutrally buoyant drop in a laminar shear flow using the IP model. Lagrangian resolution of (a) $N_\text{vertices}=640$, (b) $N_\text{vertices}=1280$. In both cases, the viscosity ratio is set to $\hat \mu=1$ (c) Comparison of the semi-axes lengths versus the Capillary number (b) Comparison of the angle formed by the major-axis in the shear plane with the velocity direction.}
\label{fig:def}
\end{figure}

To estimate the ad-hoc surface tension value for any given spring network the following steps are undertaken. We first fix the Lagrangian resolution i.e the number of vertices on the surface of a sphere and initialize a spherical drop under a given shear rate $\dot \gamma$ with a set of elastic constants. For the first set of elastic constants, $k_e$ and $k_a$ are fixed to large values in comparison to $k_b$ thus resulting in an extremely stiff drop. $k_v$ is chosen to be much larger than the rest of all constants as this ensures incompressibility of the immersed drop. Once the first set of elastic constants are chosen the drop is allowed to deform under the action of the velocity gradient $\dot \gamma$ according to the potential approach described in the previous section. If the final state of the drop is close to spherical, the elastic constant $k_e$ and area constant $k_a$ are reduced simultaneously which reduces the overall stiffness resulting in deformation of the spring network. Here it is important to note that if both $k_v$ and $k_a$ are fixed to a large value, which would imply conservation of both the volume and total area, the triangulated sphere would represent a vesicle. To represent a drop, both $k_e$ and $k_a$ are reduced to a low enough value such that the drop deforms approximately into an ellipsoid as shown in figures \ref{fig:def}(a), (b). The initial and final states of the triangulated sphere shown in figure \ref{fig:def}(a),(b) are for two different Lagrangian resolutions i.e the spheres are discretised using 320 and 1280 vertices, respectively.

 Once the deformation of the drop has reached a steady state, we compute the semi-major axis ($L$) and semi-minor axis ($B$) of the deformed sphere using the IP model. Next, we use a phenomenological model proposed by Maffettone and Minale \cite{maffettone1998equation} (hereafter called \lq MM\rq\ model) to estimate the Capillary number $Ca$ for which a neutrally buoyant immersed drop would have the same final state under similar flow conditions. The model proposed by MM \cite{maffettone1998equation} predicts the deformation of a drop in an arbitrary velocity field under the assumption that the drop is ellipsoidal in shape. For a simple flow field such as a laminar shear flow the model can be analytically solved to give the steady state values of the semi-major ($L$) and semi-minor axis ($B$) of the deformed drop as given below. 

\begin{equation}
L^2 = \frac{f_1^2+Ca^2+f_2Ca\sqrt{f_1^2+Ca^2}}{(f_1^2+Ca^2)^{1/3}(f_1^2+Ca^2-f_2^2Ca^2)^{2/3}}
\label{eqn:La2}
\end{equation}

\begin{equation}
B^2 = \frac{f_1^2+Ca^2-f_2Ca\sqrt{f_1^2+Ca^2}}{(f_1^2+Ca^2)^{1/3}(f_1^2+Ca^2-f_2^2Ca^2)^{2/3}}
\label{eqn:Ba2}
\end{equation}
In equations (\ref{eqn:La2}) and (\ref{eqn:Ba2}), $f_1$ and $f_2$ are constants which depend on the viscosity ratio ($\hat \mu$) and $Ca$ is the Capillary number.

\begin{equation}
f_1=\frac{40(\hat \mu+1)}{(2\hat \mu+3)(19\hat \mu+16)} \qquad f_2=\frac{5}{2\hat \mu+3},
\label{eqn:pfact}
\end{equation}

This model has already been used in other studies; for example to predict hemolysis of red blood cells \cite{de2012computational} and also deformation and orientation statistics of drops in turbulent flows \cite{biferale2014deformation,spandan2016deformation}. Additionally, experimental studies have shown that under moderate deformations the steady-state droplet shape can be very well described by an ellipsoid \cite{torza1972particle,guido1998three}.

In figure \ref{fig:def}(c), (d) we plot the analytical solutions (MM model - solid lines) in the form of the lengths of the semi-axes and the orientation angle of the major axis (corresponding to the axes with length $L$) with the stream-wise direction versus the Capillary number. Using this as a reference, we check the position of overlap of the semi-axes lengths computed through the IP model with the MM model to estimate the corresponding Capillary number. This match is shown through a vertical dotted line in figure \ref{fig:def}(c) and since the flow configuration such as drop radius, viscosity and shear rate are already fixed in the simulation, this Capillary number can be directly used to estimate the ad-hoc surface tension value for the chosen elastic constants. The left and right panels in figure \ref{fig:def} correspond to different Lagrangian resolutions. The important point to observe here is the reasonably good match between the semi-axes lengths computed from the IP model and the MM model. Small differences in the semi-axes lengths could arise due to multiple reasons, (i) lack of sufficient Lagrangian resolution since in the IP model the surface of the sphere is discretised using markers (ii) MM model assumes a perfectly shaped sphere which deforms into an ellipsoid while the IP model has no constraint of deforming into an ellipsoid (iii) the elastic constants would need further tuning.

Next, we keep the elastic constants the same and change the Capillary number which can be done by either changing the shear rate or the viscosity of the fluid. As shown in figure \ref{fig:def}(c,d) again the semi-axes lengths computed from the IP model agree reasonably well with the analytical solutions from the MM model. This shows that the ad-hoc surface tension computed by fitting the results from a single simulation using the IP model with MM model is reliable to extend the approach to other flow conditions. A good agreement with the MM model is found also for the orientation angle of the semi-major axes as shown in figure \ref{fig:def}(d). At higher Capillary numbers ($Ca=0.2$) there is some difference found in the lengths computed from the IP model as compared to MM model (left panel of figure \ref{fig:def}(c)). However, this is just an effect of the Lagrangian resolution and can be corrected by increasing the number of vertices on the surface of the sphere as is seen in the right panel of \ref{fig:def}(d).

\subsection{Dynamics of a liquid-liquid interface deforming in cross-flow}

\begin{figure}
  \centerline{\includegraphics[scale=1.0]{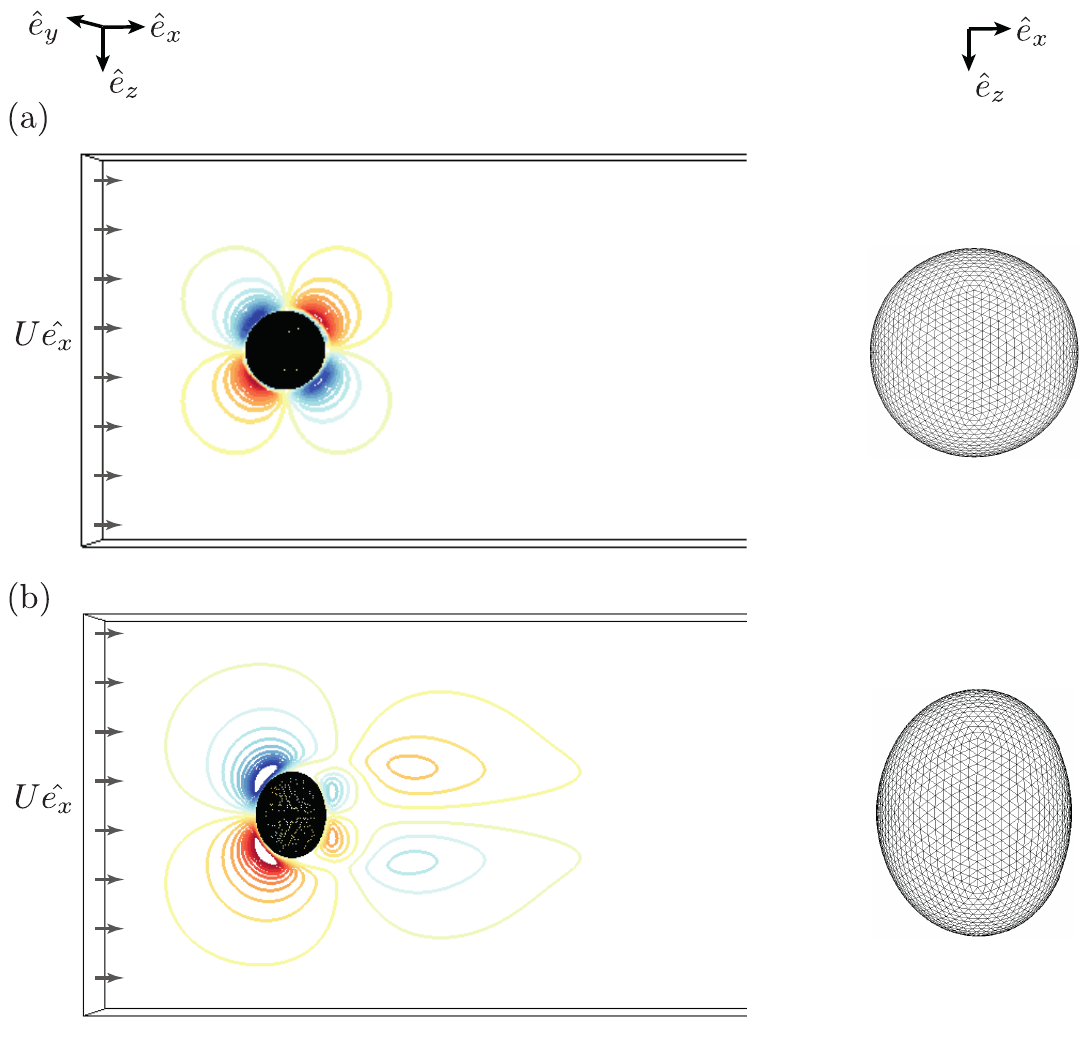}}
  \caption{Left panels show the contours of velocity in $\hat e_z$ direction along with the deforming drop at two different time instants. Right panels show the corresponding drop in the form of the deformed triangulated spring network. The Reynolds number of the flow based on the initial drop diameter is set to $Re=150$, while the elastic constants chosen correspond to a Weber number $We=2$.}
\label{fig:ifweb}
\end{figure}

\begin{figure}
  \centerline{\includegraphics[scale=1.0]{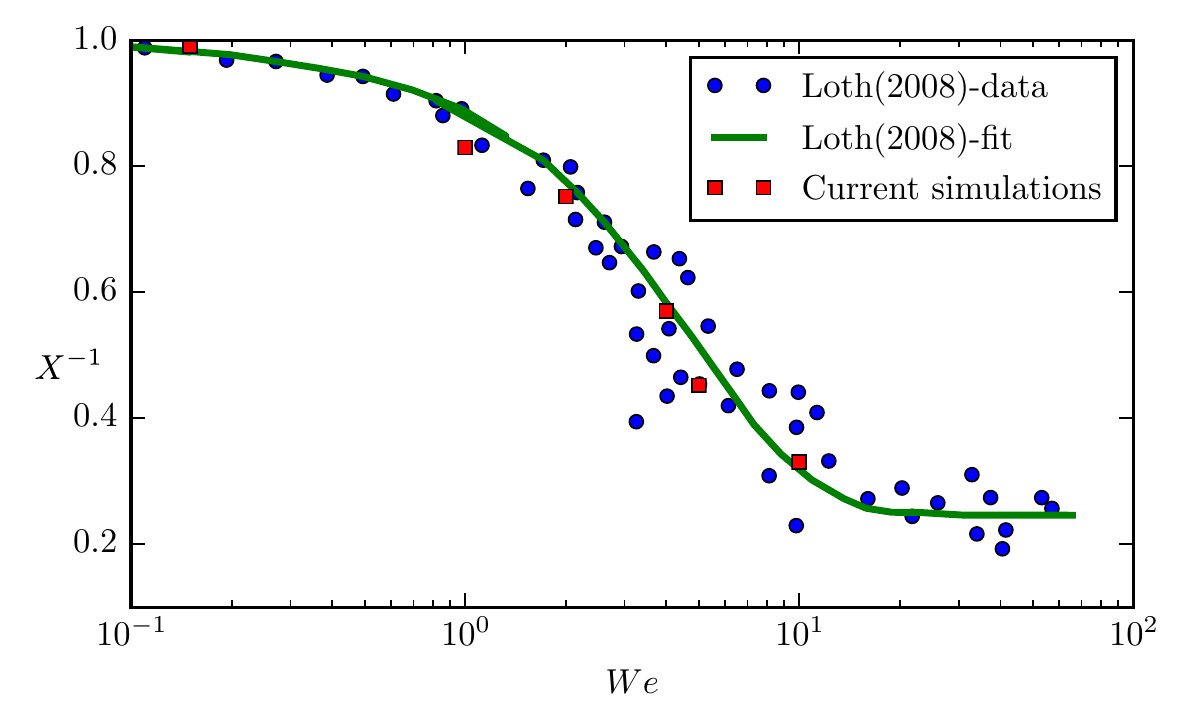}}
  \caption{Comparison of the inverse aspect ratio ($X^{-1}$) of the deformed drop versus Weber number at $Re=150$ with data from Loth \cite{loth2008quasi} for contaminated drops or bubbles}
\label{fig:lothcomp}
\end{figure}

In the previous subsection we demonstrated that by tuning the elastic constants for a single flow configuration to compute an ad-hoc surface tension, the IP model can be used to reliably simulate a neutrally buoyant drop deforming in a laminar shear flow. We now move on to simulating a more dynamic problem where the interface is strongly linked to the local flow conditions. In order to do this we take the same test case as done by Schwarz \emph{et al.} \cite{schwarz2016immersed} and have a drop immersed in a cross flow and compute the mean shape arising from the resulting flow conditions. Here, it is important to note that unlike the previous sub-section the IBM forcing is turned on i.e. $f_b \neq 0$ and is computed as described in the previous section. For such a flow, the aspect ratio of the deforming drop depends strongly on the Weber number $We=\rho_fU^2_\text{ref}d_\text{eq}/\sigma$, which is the ratio of inertia forces acting on the drop in comparison the surface tension forces (for more details see the review by Loth \cite{loth2008quasi}). The cross-stream set up in the domain is influenced by the interface of the spherical drop leading to the development of a boundary layer on the drop surface and a corresponding wake.

The computational domain is taken to be of size $L=(10,5,5)d_{\text{eq}}$, $d_{\text{eq}}$ is the diameter of the drop in its initial spherical shape. The spherical drop is triangulated with $N_v=2562$ nodes and is placed at $\pmb x = (0.5,0.5,0.5)L_z$. The vertical direction ($\hat e_z$) is wall bounded with stationary free-slip walls; $\hat e_y$ direction is periodic in nature and a uniform flow of $\pmb U = U\hat e_x$ is imposed in the $\hat e_x$ direction.

The control parameters for such a problem are the Reynolds number, $Re = Ud_{\text{eq}}/\nu_f$ and the Weber number, $We=\rho_fU^2_\text{ref}d_\text{eq}/\sigma$. The response of the system can be measured through the quantification of the wake of the drop and also through the morphology of the drop. The combined action of the dynamic pressure acting on the faces of the drop and the shear stresses generated from the boundary layer development on the surface of the drop leads to its deformation. In figure \ref{fig:ifweb} we show the wall-normal component of the velocity field ($u_z$) and the corresponding deformed drop represented through the triangulated spring network. The two snapshots shown in figure \ref{fig:ifweb} are at two different instants showing the starting up phase and the deforming phase. To quantify the shape of the immersed drop we compute the mean aspect ratio of the bubble measured as the ratio of the lengths of the drop bounding box in the wall-normal and stream-wise directions i.e. $X=l_z/l_x$, where $X$ is the aspect ratio and $l_z$, $l_x$ are the lengths of the box surrounding the deformed drop in the $\hat e_z, \hat e_x$ directions, respectively. In figure \ref{fig:lothcomp} we plot the inverse of the measured aspect ratios of the deformed drop versus the corresponding Weber number and compare it with experimental data from multiple measurements \cite{loth2008quasi}. For these simulations the Reynolds number is fixed to $Re=150$ and the Weber number is changed by modifying the elastic constants for each simulation. A very good match is found between the aspect ratios computed from the IP model and the several experimental measurements of drop shapes found in literature. These simulations further show that the IP model can be reliably used to simulate deformation in liquid-liquid interfaces under given flow conditions.

\section{Dynamics of the left heart ventricle}
We now move on to simulating the flow inside the left ventricle of the heart where the motion of the ventricle and the valves are fully coupled to the flow dynamics. The results from the numerical simulations are compared against ad-hoc experiments where the ventricle is made up of silcone rubber.

The various structures used for this simulation are shown in figure \ref{fig:lagmesh} and it is important to note that each structure is made up of a different material i.e. each material has a different elastic property. The left ventricle and the natural mitral valve can move and deform based on the local flow; the leaflets of the mechanical mitral valve, while rigid in shape, can move depending on the forces acting on their faces and more specifically on the moments of the pressure and viscous forces about the hinges of the leaflets; the channels for the aortic and mitral valves are completely rigid, fixed in space and provide a passage for the influx and outflux of the flow. The aortic valve is not simulated explicitly in these simulations but only through an opening/closing mechanism that is imposed by the immersed boundary depending on the phase of the cycle. While this has been done to limit the computational effort, it has no major consequences on the results because we are only interested in the ventricular flow and the aortic valve influences flow mainly in the ascending aorta. The dynamics of the aorta could affect the ventricular flow because of the timing of the opening and closure of the aperture, but it is driven by the impedance of the circulatory system downstream and its simulation is much more complicated and out of scope of this paper. 

In reality, the configuration of the left ventricle is determined by the dynamics of the myocardium contraction and relaxation along with the deformation of the valves and vessel walls. The complete structure adjusts to the forces induced by the hydrodynamic loads (pressure and shear stresses), body forces, internal damping and the internal elastic forces. In our simulations, the flow into the ventricle is governed through an inflow-outflow channel rather than a myocardium contraction to facilitate comparison with experiments. Similar to the experiments, the ventricle is assumed to be made of a homogeneous material i.e. silcone rubber. With minor modifications the IP approach works equally well for hyper elastic or inhomogeneous (orthotropic) materials as discussed in section \ref{sec:eqn} and by de Tullio and Pascazio \cite{detullio2016moving}. 

\subsection{Experimental and numerical setup}

\begin{figure}
  \centerline{\includegraphics[scale=1.0]{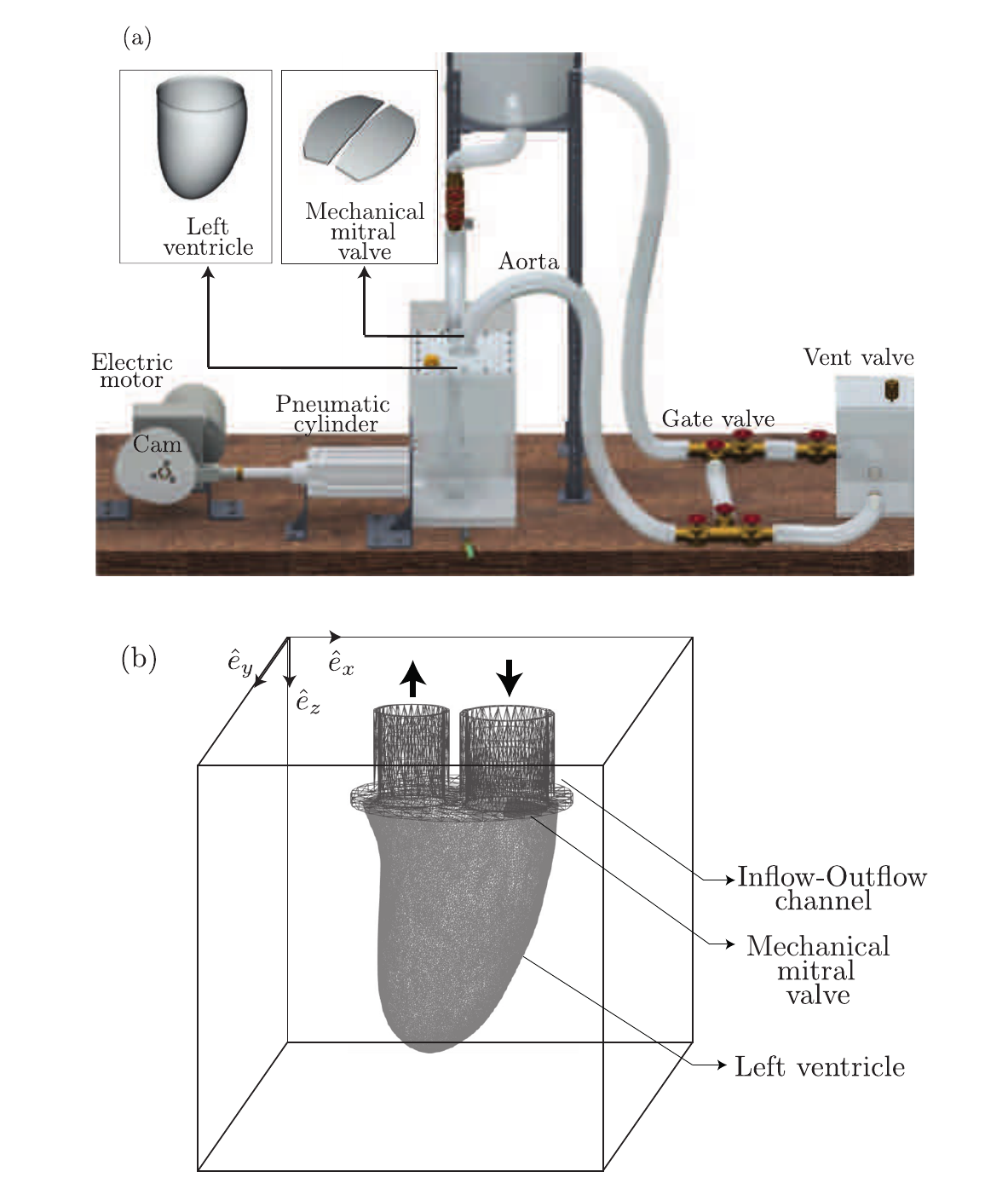}}
  \caption{(a) CAD rendering of the experimental setup built for validating the numerical approach (b) Cartesian computational box with the inflow-outflow channel, mitral valve and the left ventricle. Individual components are shown in figure \ref{fig:lagmesh}}
\label{fig:exp}
\end{figure}

\begin{figure}
  \centerline{\includegraphics[scale=1.0]{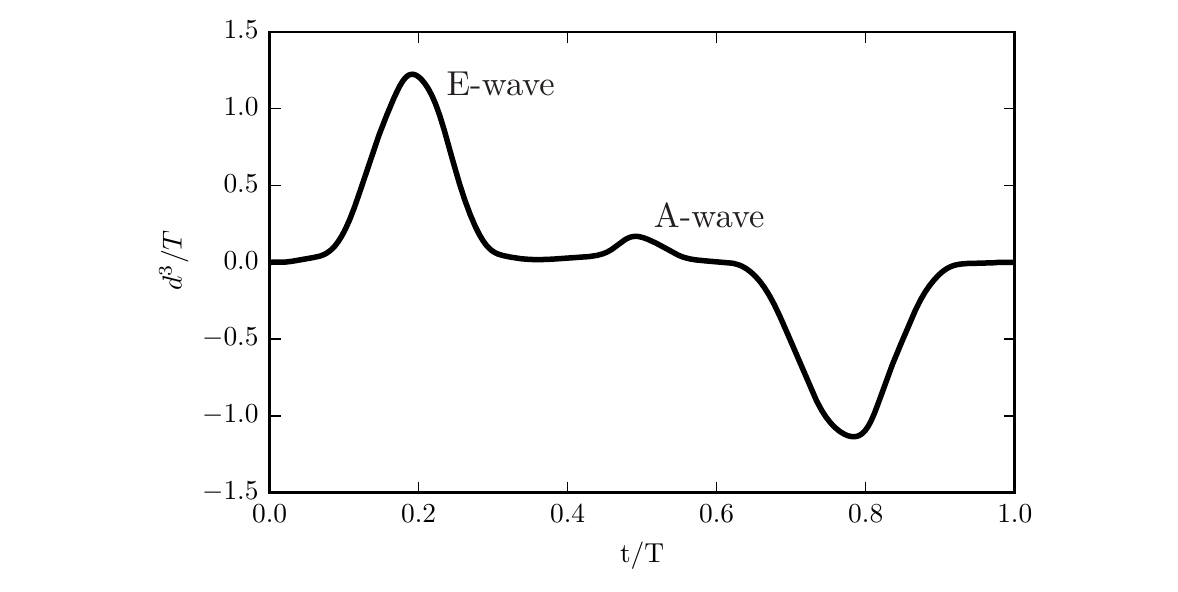}}
  \caption{Flow rate (scaled by the diameter $d$ of the opening to the mitral valve) versus time (normalised using time $T$ required for one cardiac cycle. The flow rate regulates the expansion (positive flow-rate or dyastolic phase) and relaxation (negative flow-rate or systolic phase) of the ventricle.}
\label{fig:flowr}
\end{figure}

In figure \ref{fig:exp}(a) we show a CAD rendering of the experimental apparatus used to replicate the dynamics of the left ventricle with a mechanical mitral valve and results from this will be used to validate the numerical model. An electric motor is used to drive a cam which imposes a prescribed displacement in time of the pneumatic piston/cylinder. The cylinder is directly linked to a Plexiglass box which is transparent and allows for the observation of the evolution of the left ventricle model inside. The time law imposed by the pneumatic cylinder is replicated by the fluid in the tank in which the left ventricle is immersed and is the only deformable element. Additionally, the evolution of the flow rate imposed by the motion of the cylinder is captured versus time and this is used as a boundary condition in the numerical simulations (figure 6). This is shown in figure \ref{fig:flowr} where we plot the flow rate versus time. As can be seen in figure \ref{fig:flowr}, the first part of the cycle has one strong peak (E-wave) and a secondary weak peak (A-wave) which is the result of the shape of the cam. The shape of the cam can be modified to achieve any desired flow rate profile. In the case shown here the ratio of amplitudes of A-wave to the E-wave is approximately 0.15. The profile of the cam is chosen in such a way that the flow rate resembles that of an inefficient and failing left ventricle and is generally observed in old people or heart patients. In a healthy condition, the time evolution of the flow rate versus time is similar to that shown in figure \ref{fig:flowr} but with an amplitude ratio of A-wave to E-wave of approximately 0.5. The efficiency/healthiness of the ventricle can also be quantified using ejection fraction ($EF$) which quantifies the pumping efficiency of the ventricle and is calculated as $EF=100(V_{\text{max}}-V_{\text{min}})/V_{\text{max}}$; $V_{\text{min}}$ and $V_{\text{max}}$ being the minimum and maximum values of the volume of the left ventricle, respectively during the cycle.

The left ventricle is transparent and made up of silicone rubber, fixed to the upper surface of the box by a rigid plate and consists of a mechanical mitral and aortic valves. The fluid (deionized water here) inside the left ventricle is pumped into the aorta which then flows into the hydraulic circuit composed of two branches. In one, the windkessel, there is a box connected in series to simulate the vascular capacitance while there are gate valves to regulate the impedance of the systemic circulation or to exclude one branch or another. The fluid after passing through the hydraulic circuit returns into the ventricle through the duct and a new cardiac cycle starts. In order to compare experimental measurements and numerical simulations we make use of Particle Image Velocimetry (PIV) measurements \cite{falchi2006robust} where the fluid is seeded with tracer particles (10 $\mu$m diameter pine pollen) and illuminated by a laser sheet. The motion of the particles is captured using a high-speed camera and a robust algorithm is used to compare image windows in subsequent frames and estimate the velocity field in the flow on a regular grid.

\begin{figure}
  \centerline{\includegraphics[scale=1.0]{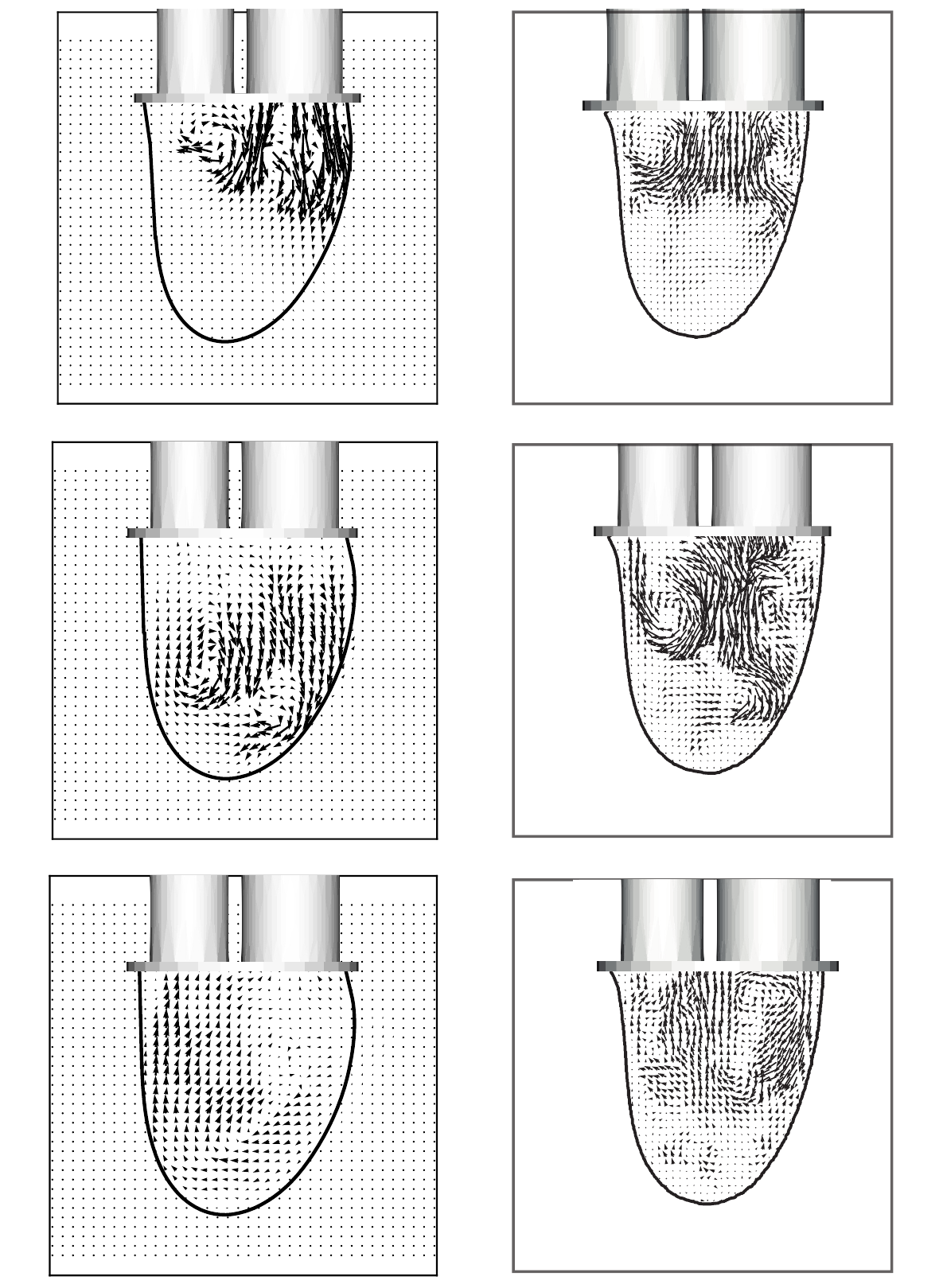}}
  \caption{Snapshots of the flow inside the ventricle during the diastolic phase in the cardiac cycle. (left panels) Numerical simulations; (right panels) Experimental measurements}
\label{fig:dyas}
\end{figure}


In figure \ref{fig:exp}(b) we show the computational domain and the setup of the complete left ventricle along with the mechanical mitral valves and the channels for the aortic and mitral valves. The geometry of the structures, the material properties and the boundary conditions have been chosen to replicate the experimental conditions as close as possible. The channels connected to the mitral and aortic valve perform the function for allowing the influx/outflux of the fluid into/from the ventricle. Since we use the IBM formulation for representing any immersed body, the whole domain is filled with a single fluid. The domain is periodic in all the directions $\hat e_x$, $\hat e_y$ and while it is confined in the $\hat e_z$ direction it allows for inflow-outflow boundary conditions on selected regions. The flow rate evolution shown in figure \ref{fig:flowr} is used as the boundary condition on the inflow/outflow channels and is linked to an amplification factor that regulates its amplitude, i.e. the higher the amplitude the higher the ejection fraction of the left ventricle. In the numerical simulations we set the value of $EF$ to 30\% which is what is imposed in the experiments in order to study the flow in a severe failing left ventricle. After performing grid independence tests, a resolution of 150x150x150 was chosen for the Eulerian field. The surface of the ventricle is discretised with 51142 triangular elements; mechanical valves with 2578 elements and the natural valves with 3794 elements each. Both experiments and the simulations are performed in dynamic similitude with a real left ventricle i.e. since the dimensions in the experiments and simulations are set to a 1:1 ratio in comparison with a real left ventricle and water is four times less viscous than blood, the total system is pulsated four times slower to maintain the same Reynolds number. The characteristic Reynolds number in the flow based on the mitral orifice diameter and maximum inflow velocity is around 5000. 

\begin{figure}
  \centerline{\includegraphics[scale=1.0]{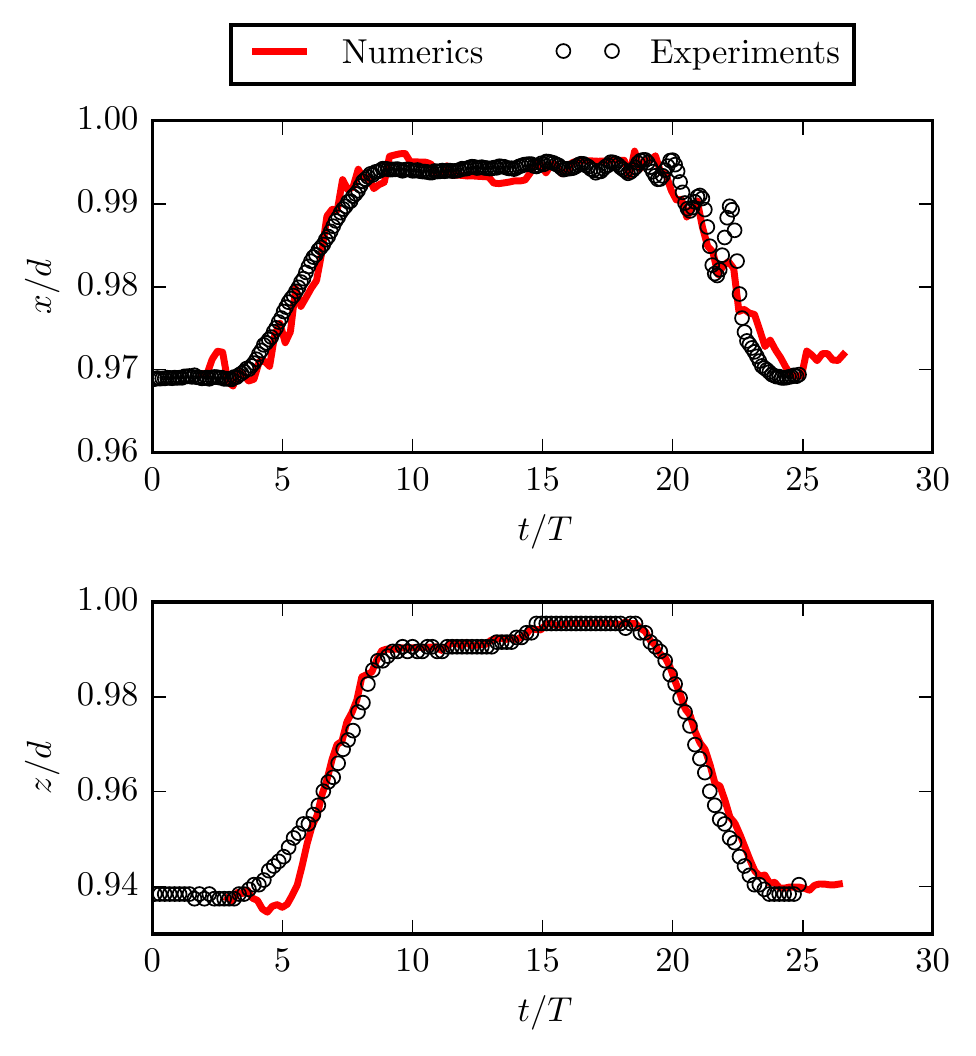}}
  \caption{Comparison of the mean position of the left ventricle in the $\hat e_x$ and $\hat e_z$ directions versus time.}
\label{fig:xzprof}
\end{figure}

We first look at the large scale flow structures created inside the ventricle. In figure \ref{fig:dyas} we plot the instantaneous snapshots of the flow velocity vectors in the mid-Y plane at certain time instants. All the left panels correspond to numerical simulations while the right panels show the measurements from the PIV experiments. It can be seen that the large scale flow dynamics can be reliably captured in the numerical simulations as compared to the experiments. During the initial part of the cardiac cycle i.e. the diastole, the jet from the mitral valve passes through the prosthetic mechanical leaflets which starts to open. The flow over the two leaflets results in the propagation of two vortices into the ventricle, one close to the left wall and the other in the centre. The two vortices are directed towards the apex of the ventricle, but since in both the simulations and experiments we reproduce the dynamics of a failing left ventricle the vortices soon dissipate into small scales and the mitral jet is not able to penetrate down to the apex and wash out the stagnant fluid. This is shown more clearly later.

We now compare the mean position of the left ventricle in the $\hat e_x$ and $\hat e_z$ directions to further validate the dynamics of the deforming ventricle from the numerical simulations. This is shown in figure \ref{fig:xzprof} where the mean position $x/d$ and $z/d$ is plotted against time. The position obtained from the numerical simulations have reasonably good agreement with its experimental counterpart except from small oscillations which cannot be captured in the experiments. This shows that not only the large scale flow structures, but also the dynamics of the deforming left ventricle which is modelled using the interaction potential approach can be simulated with reasonable accuracy.

\begin{figure}
  \centerline{\includegraphics[scale=1.0]{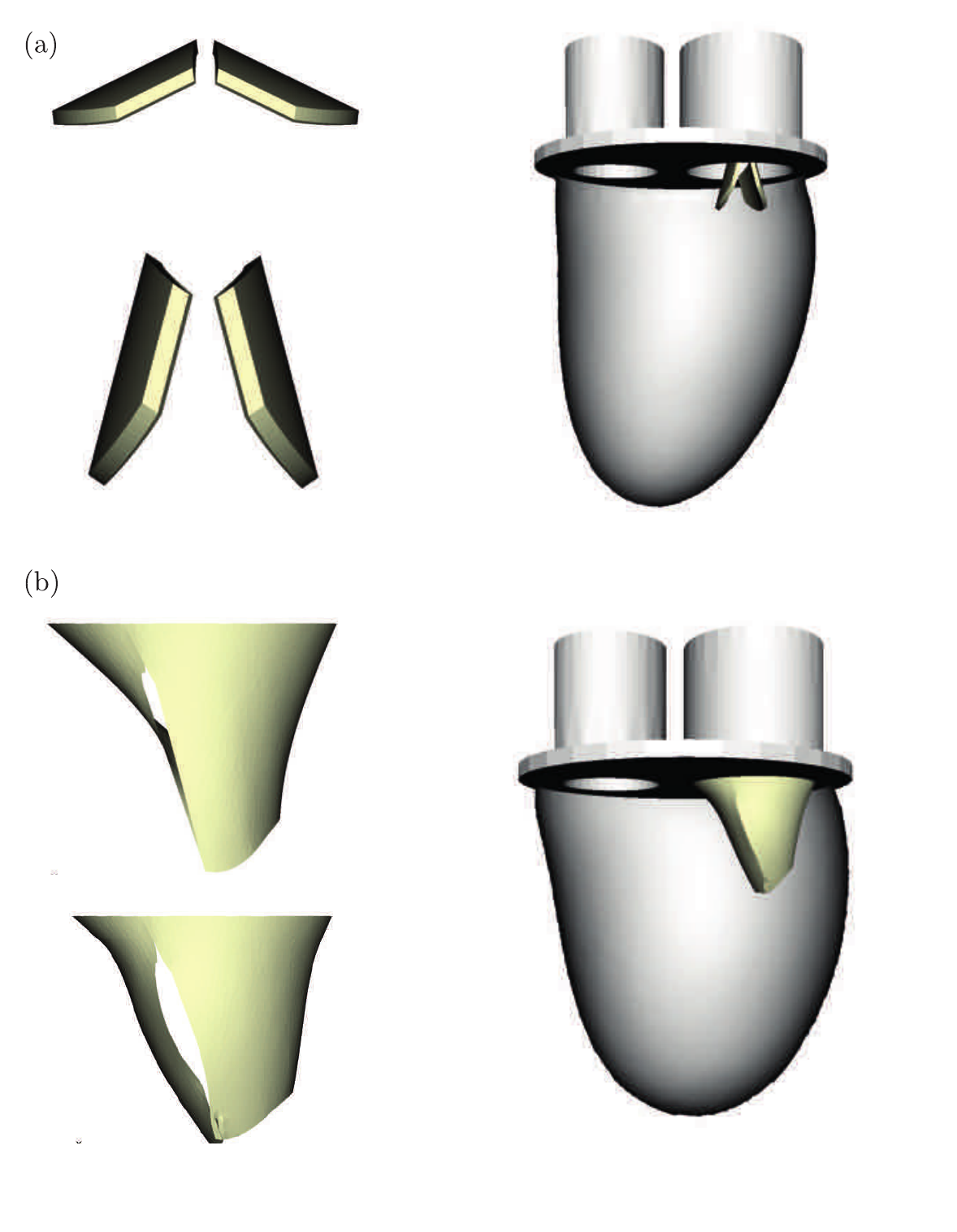}}
  \caption{Left panels show snapshots of (a) Mechanical and (b) Natural valves at two different time instants in the diastolic phase of the cardiac cycle. The right panels show the full set up of the ventricle along with the valves and the inflow/outflow channels.}
\label{fig:valvecomb}
\end{figure}

\begin{figure}
  \centerline{\includegraphics[scale=1.0]{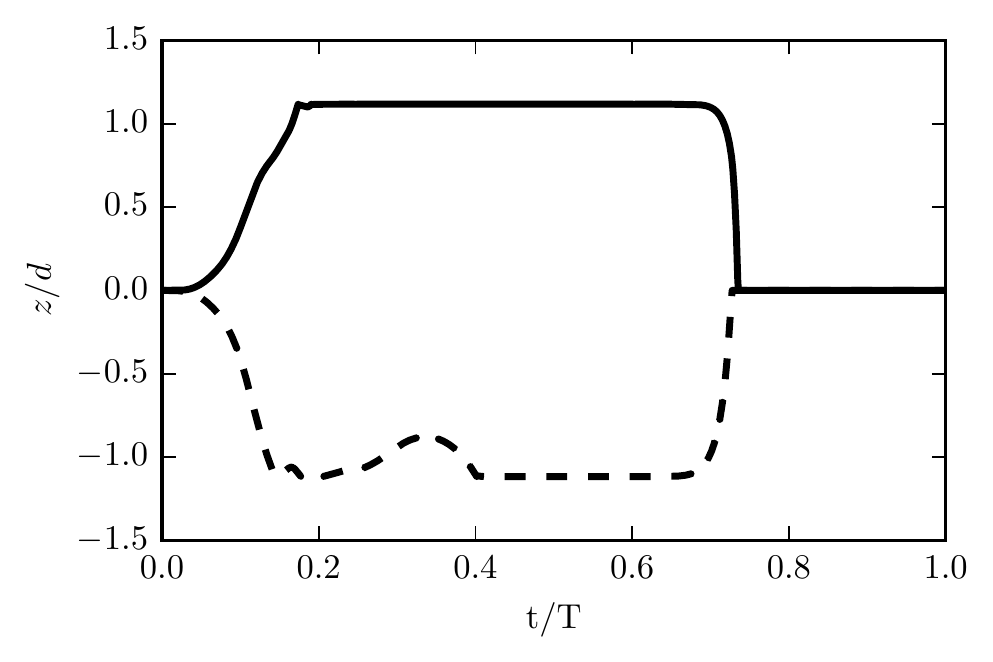}}
  \caption{Axial position of the centre of mass of prosthetic mechanical leaflets for an healthy left ventricle in a single cardiac cycle. The solid line and dashed line represent two different leaflets. $d$ is the diameter of the mitral orifice, while $T$ is the time taken for one full cardiac cycle.}
\label{fig:leaf}
\end{figure}

\begin{figure}
  \centerline{\includegraphics[scale=1.0]{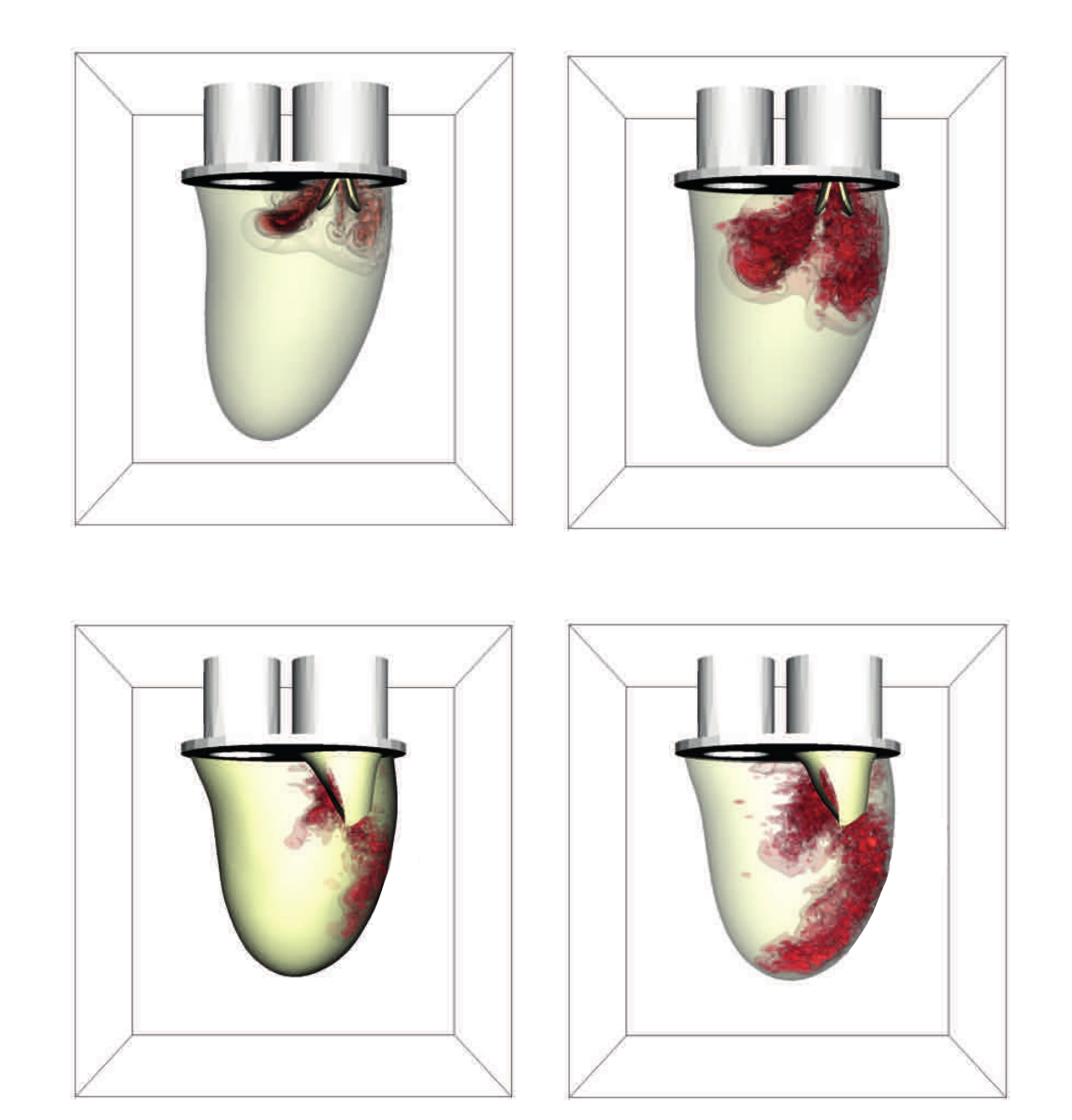}}
  \caption{Snapshots of the flow inside the ventricle at the two different time instants (left and right panels) in the diastolic phase of the cardiac cycle. Top panels show the ventricle with a mechanical natural valve while the bottom panels show the ventricle with a deformable natural valve. The colour represents the iso-surface of the velocity magnitude of the flow inside. }
\label{fig:heartvol}
\end{figure}

\subsection{Mechanical and natural mitral valve}

An important element affecting the dynamics and nature of the flow is the presence of the mitral valve. To this purpose we couple our computational model of the left ventricle with two kinds of mitral valve (i) prosthetic mechanical valve (ii) natural mitral valve. While the mechanical valve is structurally rigid the natural valve is similar to a flexible membrane and can deform based on the local flow conditions (c.f. figure \ref{fig:lagmesh}). In figure \ref{fig:valvecomb} we show both the mechanical and natural valve during their initial state and when they are close to being fully open. Here we would like to emphasize that since the numerical set up is a full fluid-structure interaction approach, the valve dynamics are solely determined by the hydrodynamic loads and any geometrical constraints set up by the user. The panels on the right in figure \ref{fig:valvecomb} show a clear difference in the shape of the ventricle. The shape of the ventricle depends heavily on the hydrodynamic loads exerted on it from the fluid inside it. The mechanical and natural mitral valves lead to different flow structures inside the ventricle and thus a different shape of the ventricle. We now show the difference in flow structures arising from the different valves used. 

First, we consider the case of a prosthetic mechanical mitral valve which in a sense obstructs the flow through the mitral orifice. For the dynamics of the full valve, we allow each leaflet to rotate around a fixed axis which is symmetric about a plane situated in the centre of the mitral orifice. In figure \ref{fig:leaf} we show the dynamics of the leaflets which go in opposite directions and close asymmetrically since the backward flow induced by the systole comes from different regions of the ventricle for the anterior and posterior leaflets. The opening phase starts at the beginning of the diastole as the flow starts accelerating and finishes before the end of the flow acceleration when the fully open position is reached. The closing phase starts when the flow rate reaches its peak and ends when the minimum negative value of the flow rate function is achieved, thus positioning the leaflets in the fully closed position. 

In figure \ref{fig:heartvol} we show instantaneous velocity fields during the diastolic phase of the cardiac cycle with both a prosthetic mechanical and natural mitral valve. In the case of mechanical valves (top panels of figure \ref{fig:heartvol}), the leaflets start rotating during the early opening phase and destabilise the mitral jet. In the bottom panels of figure \ref{fig:heartvol} we show the flow structure in the presence of a natural mitral valve which is also made up of two leaflets but has different dynamics due to the inherent deformability of the natural valves. In the presence of the mechanical valve, the flow is split into three different jets thus causing high vorticity regions in the wake of the valve. This results in the mitral jet not reaching the bottom of the ventricle as desired. It is evident that the disturbance generated by the mechanical leaflets destabilizes the mitral jet, creating vortex rings thus further decreasing its capability to penetrate the ventricular region. The flow soon degenerates into small scales that are dissipated during the diastatic phase of the cycle. Unlike the prosthetic mechanical valves, the natural valves can deform based on the local hydrodynamics forces allowing for a much smoother flow of the mitral jet into the ventricle. Due to this the natural valves evolve differently resulting in a different flow structure in the ventricle which reaches the bottom of the ventricle which is a desired flow condition.

From the discussion of figure \ref{fig:heartvol} it is clear that the behaviour of the flow inside the left ventricle depends strongly on the kind of mitral valve used.  Overall, we have been able to show that the complete dynamics of the left heart ventricle with either mechanical or biological valves can be simulated reliably using IBM coupled with an interaction potential approach for deformation.

\section{Data Structure and Parallelisation Strategy}

In this section we describe the parallelisation strategy implemented and the data structures required for the simulations described in the previous sections. For parallelising an IBM code which deals with a suspension of spherical particles, Uhlmann \cite{uhlmann2004simulation} proposed a \lq master\rq\ and \lq slave\rq\ strategy, where each particle is allocated an individual \lq master\rq\ processor which is responsible for all the computations related to it. Additional \lq slave\rq\ processors may be allocated to help the \lq master\rq\ processor. Wang et al. \cite{wang2013parallel} employ a \lq gathering-scattering\rq\ strategy where a single master processor is responsible for the computation of the Lagrangian force on the immersed bodies and advecting them and this information is scattered to the slave processors which solve the Navier-Stokes equations in parallel. While both parallelisation approaches have been shown to produce reasonable performances, there exist some drawbacks and challenges. The strategy implemented by Uhlmann \cite{uhlmann2004simulation} requires continuous exchange of control on the Lagrangian mesh by the processors which may lead to a complex programming environment. The approach of Wang et al. \cite{wang2013parallel} eliminates this issue leading to a simple structure of the code, increase in the memory usage on the master processor and data transfer between the master and slaves are some hurdles. In this work, we propose a different parallelisation approach for the IBM where the information of all triangle nodes is present with all processors, while the computation required for each Lagrangian node/structure is performed only by specific processors depending on the type of computation that needs to be performed. In other words the allocation of processor for the IBM depends on the task that needs to be performed which results in a task-based parallelism for the FSI-IBM computation. 

We will first describe in brief the parallelisation strategy employed for the flow solver and later explain the data structures and parallelisation implemented for the FSI-IBM. For the flow solver we employ domain decomposition and split the Cartesian box into slabs (\lq one-dimensional slab\rq\ parallelisation). In addition to the slabs, each processor needs to store information from the neighbouring processors which would be required for computing the derivatives and is stored in what is called as a \lq halo/ghost\rq\ layer. Since the flow solver employs a second-order finite difference spatial discretisation at most one halo layer is required on each side of a slab for single phase flows. However, as we explain later when this solver is coupled with a FSI-IBM solver for finite-size bodies which makes use of MLS interpolations, multiple halo layers become necessary. It is important to keep in mind that an unrestricted increase in the number of stored halo layers would automatically result in an increase in the communication time which may deteriorate the overall performance of the code. For the MLS interpolations which need a support domain of 27 (3x3x3) Eulerian points at most 3 halo layers are necessary.

The Lagrangian meshes shown in figure \ref{fig:lagmesh} are unstructured and are exported in the form of a GTS (GNU Triangulated Surface) data format which contains information about the spatial positions of the vertices of the triangular elements, the various vertices which are connected by edges and also the edges which constitute a face. Using this information we construct additional auxiliary arrays which will be required while computing the total force acting on the triangle nodes based on the potentials described in the section \ref{sec:eqn}. The total number of vertices, edges and faces on a single immersed body is stored in \texttt{N\_vert}, \texttt{N\_edge}, \texttt{N\_face}, respectively while \texttt{N\_particle} is the total number of immersed bodies to be simulated and \texttt{N\_edge\_vert} is the maximum number of edges that any single vertex can be connected to. A brief overview of the required auxiliary integer arrays is given below.

\begin{enumerate}
\item \texttt{vert\_of\_edge[2, N\_edge, N\_particle]} : Contains pairs of vertices sharing a single edge.
\item \texttt{face\_of\_edge[2, N\_edge, N\_particle]} : Contains pairs of faces sharing a single edge.
\item \texttt{vert\_of\_face[3, N\_face, N\_particle]} : Contains the three vertices that constitute a single face.
\item \texttt{edge\_of\_face[3, N\_face, N\_particle]} : Contains the three edges that constitute a single face.
\item \texttt{vert\_of\_vert[N\_edge\_vert, N\_vert, N\_particle]} : Contains all the vertices that a single vertex is connected to.
\item \texttt{edge\_of\_vert[N\_edge\_vert, N\_vert, N\_particle]} : Contains all the edges that a single vertex is connected to.
\item \texttt{v1234[4, N\_edge, N\_particle]} : Contains all the four vertices that is contained in two faces sharing an edge.
\item \texttt{pind[3, N\_face, N\_particle]} : Stores the [\texttt{N\_x,N\_y,N\_z}] indices of each centroid relative to the Eulerian mesh and tells us inside which Eulerian computational cell the centroid resides in. This array is updated every time step.
\item \texttt{bboxind[6,N\_particle]} : Stores the indices of the bounding box of each immersed body.
\end{enumerate}

\begin{figure}
  \centerline{\includegraphics[scale=1.0]{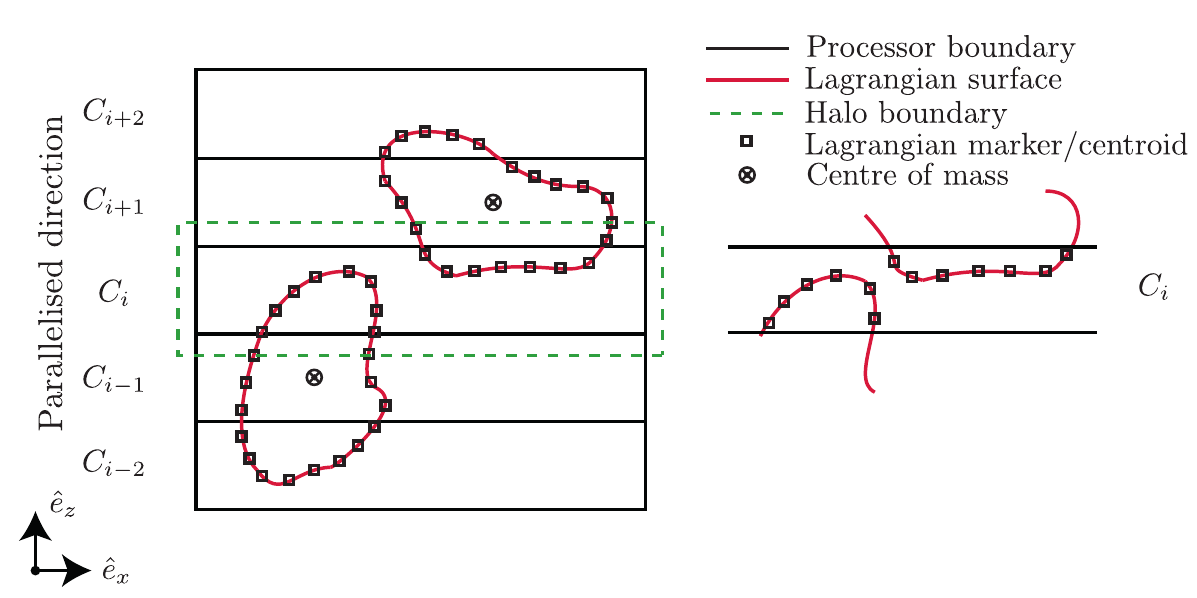}}
  \caption{Schematic of two bodies immersed in a flow. Flow solver is 1D slab parallelised. $\hat e_z$ is the wall-normal direction. }
\label{fig:slab}
\end{figure}

Consider two arbitrarily shaped deformable bodies immersed in a flow as shown in figure \ref{fig:slab}, where the squares represent the Lagrangian markers/centroids of the triangular elements. The allocation of Eulerian slabs to each processor is straight forward as the Eulerian mesh stays fixed in time and this is done at the start of the simulation. For the flow solver each processor with identity \texttt{myid} is allocated the task of solving equation \ref{eqn:ns} on a slab of \texttt{[1:N\_1,1:N\_2,N\_3\_start:N\_3\_end]}. Given below are the steps undertaken to complete one full time step of the simulation. As given below there are four major steps and multiple sub-steps involved in completing one full iteration. The steps shown here are applicable for a loosely coupled approach which has been used for the simulations in this paper; details on the strongly coupled approach are elaborated in the paper by de Tullio and Pascazio \cite{detullio2016moving}.

\begin{enumerate}
\item Compute the indices of all markers/centroids on the Lagrangian mesh relative to the Eulerian mesh.
\item Compute the properties of the Lagrangian mesh, i.e. surface areas and normals of each face of the Lagrangian mesh.
\item Compute flow configuration along with IBM forcing i.e. all three sub-steps of RK3 integration.
	\begin{enumerate}
	\item Compute intermediate fluid velocity under the RK3 framework.
	\item Interpolate velocity on the centroids of the Lagrangian mesh using MLS interpolation.
	\item Communicate the forces in the halo cells to neighbouring processors.
	\item Correct intermediate velocity using the MLS-interpolated force.
	\item Solve pressure correction equation and compute the pressure and solenoidal velocity field.
	\end{enumerate}
\item Compute external and internal loads on the immersed body.
	\begin{enumerate}
	\item Compute the external loads which is the sum of pressure and viscous forces on each face using MLS 				interpolation.
	\item Sum up external loads on all faces across all processors.
	\item Compute internal loads which are derived from the potentials described in section \ref{sec:eqn}.
	\item Sum up internal loads across all processors.
	\item Update the nodes of the triangles using Newton's law of motion.
	\end{enumerate}
\end{enumerate}

In the first step, we compute the indices of all the centroids on every triangular element and store it in a global array \texttt{pind}. In addition to the axial index of every triangular element we also compute the mean axial index of every immersed body i.e for an immersed body \texttt{i} the mean axial index is \texttt{bboxind[1:3,i]=0.5*(max(pind[1:3,:,i]+min(pind[1:3,:,i]))}. In step 2, we compute the geometrical properties of the triangulated mesh (i.e. surface areas and normals of each triangular element). Both steps (1 and 2) are done by all processors (i.e. \texttt{MPI\_COMM\_WORLD}) on all immersed bodies and at the end of this operation every processor has information on all three indices [\texttt{pind(1:3,N\_face,N\_particle)}] of every centroid immersed in the flow, surface areas and normals of every triangular element.

For steps 3(a), 3(d) and 3(e) each processor performs all the operations required on its respective slabs. Step 3(b), which consists of interpolation using MLS and computing the IBM force has to be performed on the Lagrangian markers (centroids here) and this is done only on the centroids lying within the processors slab (c.f. right panel of figure \ref{fig:slab}). This allocation is regardless of which immersed body it belongs to. This is achieved by first performing a check on the axial index of every centroid (stored in \texttt{pind[3,:,:]} and computed in step 1); for example, if the processor $C_i$ is responsible for the slab \texttt{[1:N\_1,1:N\_2,N\_3\_start:N\_3\_end]} the following procedure is undertaken.

\begin{verbatim}
do i=1,N_particle
  do j=1,N_face

    if pind(3,j,i) >= N_start(myid).AND.pind(3,j,i) < N_end(myid)
        - Perform MLS interpolation around the centroid.
        - Compute IBM forcing.
    end if

  end do
end do
\end{verbatim}

As explained in section \ref{sec:eqn}, MLS interpolations require a support domain built from 3 Eulerian grid nodes in each direction. Thus the forcing computed from a centroid lying right next to a processor boundary would be stored in a halo layer and this is communicated to the neighbouring processors in step 3(c). Every processor adds the IBM forcing received from the halo cells of the neighbouring processors to the already existing forcing thus accounting for the forcing from the centroids lying on processor boundaries.

Step 4 involves computing the external forces on the immersed body (i.e. pressure and viscous forces) which are performed following the procedure described in section \ref{sec:eqn}. The allocation of processor for computing the external processors is done in a similar manner to step 3(b). Here it is important to note that for centroids lying on the processor boundaries the probes may lie in the neighbouring processor. For example, a centroid belonging to processor $C_i$ may have an axial index of \texttt{N\_3\_start} and the axial index of the corresponding probe would be \texttt{N\_3\_start-1}. Building a support domain around \texttt{N\_3\_start-1} would require information from [\texttt{N\_3\_start-2, N\_3\_start-1, N\_3\_start}] i.e. at least two halo layers need to be stored by each processor.

\begin{verbatim}
do i=1,N_particle
  do j=1,N_face

    if pind(3,j,i) >= N_start(myid).AND.pind(3,j,i) < N_end(myid)
        - Compute probe and build support domain around the probe.
        - Perform MLS interpolation around the probe.
        - Compute pressure and viscous forces on faces.
        - Distribute the forces from faces to nodes.
    end if

  end do
end do
\end{verbatim}

In step 4(b), we reduce the external forces ($\pmb F_{\text{ext}}$) over all the triangle nodes immersed in the flow. \texttt{MPI\_ALLREDUCE} is used to perform this operation which results in all the processors having information on the external forces acting on all triangular nodes. The total force acting on each triangular node is computed as the summation of the external forces (pressure + viscous) and the internal forces arising from the elastic potentials i.e. $\pmb F=\pmb F_{\text{ext}}+\pmb F_{\text{int}}$; for the first time step the immersed body is in its reference state and all internal forces are equal to zero and in every succeeding time step $F_\text{int}$ is the internal elastic forces computed in the previous time step. With this step every processor updates the position of the triangle nodes based on the total force.

Step 4(d) involves computing the internal elastic forces derived from the potentials on each immersed body. Since this requires the full body to be treated as a whole, we compute the location of the mean axial index of every individual immersed body from the information in the array \texttt{pind}. The processor responsible for this axial index takes care of computing all the internal elastic forces (i.e. in-plane deformation, out-of-plane deformation, volume constraint and area constraint) and computing the net internal force acting on each node belonging to its allocated immersed body. While computing the internal forces on each immersed body does not require any information from the Eulerian mesh, such an allocation ensures the computing load is distributed evenly across all processors. Also it is important to note that MLS interpolations which are the computationally expensive steps in this FSI-IBM code are still performed only by processors containing the Lagrangian markers.  The pseudo code for this operation is given below.

\begin{verbatim}
do i=1,N_particle

    if z_ave(i) >= N_start(myid).AND.z_ave(i) < N_end(myid)
        - Compute forces from in-plane deformation.
        - Compute forces from out-of-plane deformation.
        - Compute forces from volume potential.
        - Compute forces from area potential.
        - Sum up forces from all potential on the nodes.
    end if

end do
\end{verbatim}

In step 4(d), we reduce the internal forces ($\pmb F^{\text{int}}$) over all triangle nodes with an \texttt{MPI\_ALLREDUCE} operation similar to the operation in 4(b). With this we complete all the steps required for one full iteration of the flow solver and the IBM coupled with the deformation.

\subsection{Scaling performance}

\begin{figure}
  \centerline{\includegraphics[scale=1.0]{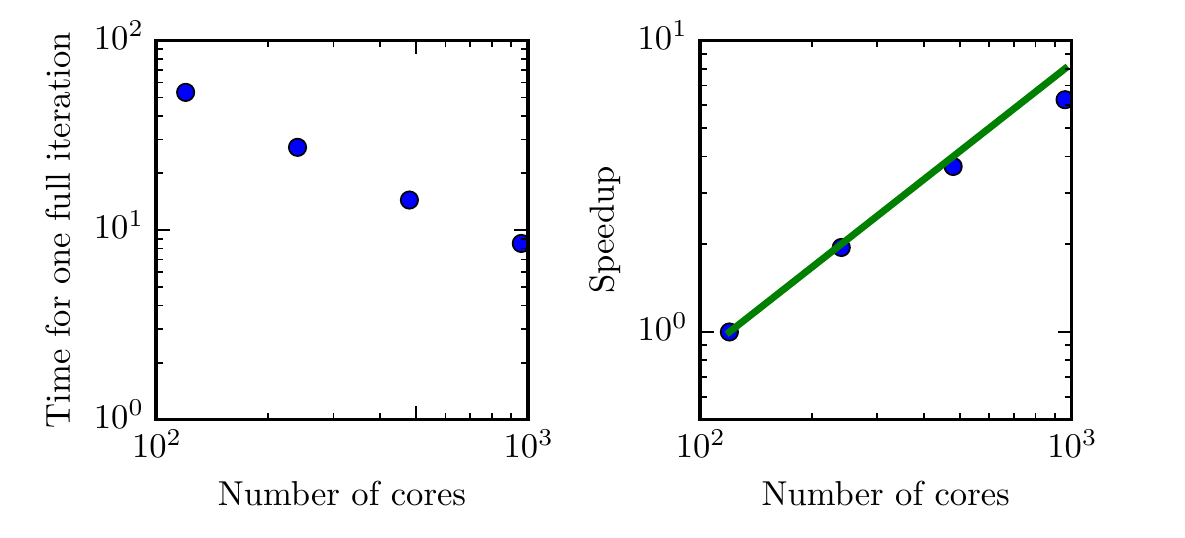}}
  \caption{(a) Scaling plot showing the time step for one full iteration of the solver versus the number of cores used. (b) Corresponding speed up versus the number of cores.}
\label{fig:scale}
\end{figure}

In figure \ref{fig:scale}, we show the computational performance of the previously discussed parallelisation strategy. These simulations were performed on the thin nodes of the Dutch supercomputing facility 'Cartesius' where each node is composed of 2x12 core 2.6 GHz Intel Xeon E5-2690 v3 CPU's. As can be seen from the plots in figure \ref{fig:scale} strong scaling is achieved up to 1000 cores. This simulation was performed on a grid of 720x720x3840 with a total of 25000 spherical particles each discretised using 320 faces, i.e. a total of 8 Million Lagrangian markers were simulated simultaneously.

As discussed in section \ref{eqn:ns}, the costliest steps in the FSI-IBM part are the ones involving MLS interpolation since each interpolation requires the construction of multiple coefficient matrices and a subsequent inversion of a 4x4 matrix (in 3D). For each Lagrangian marker (centroid) immersed in the flow two MLS interpolations are required; one at the Lagrangian marker itself to compute the IBM forcing and another at the position of the probe projected from the centroid which is used for the computing the value pressure and velocity gradients. Additional matrix operations are required for the velocity gradients since instead of the shape function we need to compute the derivative of the shape function \cite{liu2005introduction}. On a single processor, increasing the number of Lagrangian markers by two times results in a three fold increase in the simulation time. It is thus crucial to see how the parallel code performs with increase in the total number of Lagrangian markers or triangular faces.

\begin{figure}
  \centerline{\includegraphics[scale=1.0]{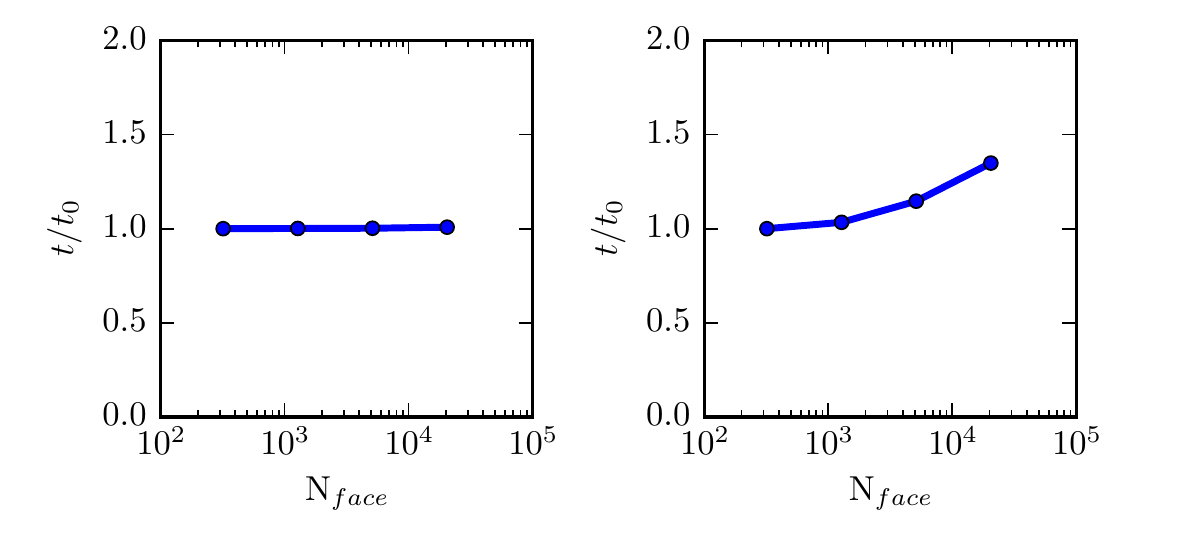}}
  \caption{Time taken for one full iteration normalised using the time taken for the first data point versus the total number of triangular elements or faces immersed in the domain. In the left panel the immersed bodies are stationary and fixed in shape while in the right panel the bodies can both move and deform.}
\label{fig:face}
\end{figure}

In figure \ref{fig:face} we plot the non-dimensional time taken for one full iteration with increasing total number of faces for two different types of simulations. The time is normalised using the time taken for the first data point i.e. $N_\text{face}=320$. For these simulations the Eulerian grid is kept fixed to 120x120x720 and a total of 120 cores were used. In the left panel of figure \ref{fig:face}, the immersed bodies are kept fixed in position and shape i.e. the computation of the structural solver is fully eliminated. Such simulations are useful to compute the hydrodynamic forces acting on stationary bodies with an mean flow imposed in the domain. As can be seen, with increase in the number of faces there is negligible increase in the computational time. In comparison to simulations involving moving and deforming objects, IBM simulations with stationary and fixed bodies require only one MLS interpolation and this is the reason for the negligible time increase. On the right panel we show the increase in time for simulations involving both moving and deforming bodies. These simulations require an additional MLS interpolation at the probe and also the computation of shape function derivatives. For such simulations, a 100 times increase in the number of faces results in approximately 1.3 times increase in the computational time. This shows that the increasing cost of MLS interpolations on the Lagrangian markers can be offset by parallelising the task over multiple processors. 

\section{Summary and Outlook}
In this paper we have demonstrated the implementation of a finite-difference based flow solver capable of handling multiple deforming immersed bodies with full fluid-structure interaction. A multi-physics interaction potential approach is used for simulating the deformation dynamics of liquid-liquid interfaces and elastic membranes. An immersed boundary method based on moving least squares interpolation is used to enforce the interfacial boundary condition on the underlying flow. In the case of liquid-liquid interfaces with the use of ad-hoc elastic constants, we have shown that the potential approach can be used to capture the deformation dynamics of neutrally buoyant drops with contaminated interfaces. By comparing with already existing analytical solutions and experimental measurements of standard configurations (deforming drop in a shear flow and cross flow), we have shown that such an approach can be reliable and self-consistent. In the second part of the paper, we have shown that the same potential approach with minor modifications to the governing equations can be used to successfully simulate the complete dynamics of the left heart ventricle with a mechanical or biological valve. In contrast to previous studies where the motion of the ventricle or the valves are imposed a priori through kinematic models, in this paper a full fluid-structure interaction simulation of the left heart ventricle was carried out on a single computing processor. The results from the simulations have been validated with ad-hoc in-house experiments. While the interaction potential approach for deformation is computationally inexpensive, parallelisation is a necessary step to simulate large scale turbulent flows with several thousands of simultaneously deforming bodies. To this effect, in the last part of the paper, we present a parallelisation strategy which was implemented for such a solver.

An exciting feature of the solver presented in this work is the computationally inexpensive and versatile nature of the algorithms used. In particular, the interaction potential approach combined with an immersed boundary based flow solver can be used to study a large class of cardiac hemodynamics problems with no pre-determined valve or ventricle dynamics. Additionally, the approach can be extended to study multiphase flows involving deformation of thousands of deforming drops and bubbles in highly turbulent flows. In our future work we are focussing on eliminating some drawbacks of the IBM used here and improving further the speed of such simulations. For example, wall-bounded turbulent flows need enhanced resolution in the Eulerian mesh near the walls to capture the boundary layers which might add a severe constraint on the Lagrangian resolution. In order to tackle this we are currently implementing a fast moving least squares implementation with adaptive Lagrangian mesh refinement to decouple the Lagrangian mesh into one for IBM forcing and another for deformation. Novel algorithms are also required to implement collision dynamics between the immersed bodies which will become crucial when several thousands of such bodies are interacting within a flow.

This work was supported by the Netherlands Center for Multiscale Catalytic Energy Conversion (MCEC), an NWO Gravitation programme funded by the Ministry of Education, Culture and Science of the government of the Netherlands and the FOM-CSER program. We acknowledge PRACE for awarding us access to FERMI based in Italy at CINECA under PRACE project number 2015133124 and NWO for granting us computational time on Cartesius cluster from the Dutch Supercomputing Consortium SURFsara. This work has also been  partially funded within the PRIN project number 2012HMR7CF by Italian Ministry of University and Research.

\section*{References}
\bibliography{mylit}

\end{document}